# Investigation of injector-coupled combustion dynamics in a methane-oxygen combustor using large eddy simulation and dynamic mode decomposition

Abhishek Sharma[1,2], Ashoke De[2,3*], Sunil Kumar[1]

[1]*Liquid Propulsion Systems Center, ISRO, Valiamala, 695547, Thiruvananthapuram, India*

[2]*Department of Aerospace Engineering, Indian Institute of Technology Kanpur, 208016, Kanpur, India*

[3]*Department of Sustainable Energy Engineering, Indian Institute of Technology Kanpur, 208016, Kanpur, India*

This paper uses a reactive flow large eddy simulation (LES) and decomposition techniques to study combustion instabilities in a methane-oxygen combustor. This work examines two case scenarios to elucidate the significance of injector-chamber frequency coupling as the cause of thermo-acoustic instability. Initial investigation in a well-known benchmark case of the continuously variable resonance combustor (CVRC) reports the potential instability mechanisms and the role of injector-chamber frequency coupling in thermo-acoustic instability. Subsequently, the multi-element rocket combustor case study identifies the critical resonant modes and highlights potential frequency coupling between the injector and the chamber region. The interplay between longitudinal pressure oscillations in the oxidizer post and transverse pressure waves in the chamber is responsible for the enhanced pressure dynamics in the combustor. The present work uses the dynamic mode decomposition (DMD) technique to reveal the evolution of acoustic modes in injector and chamber for CVRC and multi-element combustor. The dominant pressure mode forms found by DMD analysis also showcase the role of injector-chamber frequency coupling in amplified combustion dynamics. The results demonstrate how the predominant cause of combustion instability in rocket combustors can be effectively determined using the high-fidelity LES framework in conjunction with the modal decomposition technique.

---

Author to whom correspondence should be addressed. Electronic mail: ashoke@iitk.ac.in

# I. INTRODUCTION

The combination of irregular heat release and the acoustic modes in the combustion chamber leads to combustion instability. Fluctuations are enhanced when heat release aligns with the sound pressure, a phenomenon first reported by Lord Rayleigh[1]. Engine failure may occur due to these instabilities, which can place excessive strain and heat fluxes on the injectors and chamber wall. Due to the many phenomena involved, including turbulence, chemical reactions, and acoustics, it can be difficult to predict combustion instabilities. Refinement of the design process and safe operation of high-pressure rocket engine systems depend heavily on understanding combustion instability[2–5]. Compared to single injector setups, the complexity of the combustion instability mechanism is significantly increased in multi-element combustion chambers. A significant issue for experimental research is analyzing the complex interactions between injectors, acoustic waves, manifolds, and feed systems[6]. Reactive flow simulations with high fidelity are highly effective in simulating large-scale unstable processes and offer valuable insights into the mechanics underlying combustion instability[7–10]. Large Eddy Simulation (LES) is crucial to accurately capture the underlying physics of combustion dynamics. The study of turbulent mixing, unsteady heat release, acoustic wave propagation, and the complex relationships between flame, acoustics, and geometry are all included in LES, which substantially contributes to our understanding of combustion instability. In sub-scale rocket experiments, LES has effectively reproduced combustion instabilities[11–14]. Many numerical investigations have been devoted to elucidating the underlying physics of combustion instability. Because of the large amount of computational power available, LES has become extremely important in rocket engines[15]. It is a perfect technique for simulating combustion dynamics, as it can represent complex unsteady flows, heat release, and acoustic wave propagation in rocket scale combustors. Purdue University has conducted experimental studies of combustion instability at operating conditions relevant to rocket engines for over a decade. These endeavors aimed to generate necessary data and create Large Eddy Simulation (LES) models that could predict thermo-acoustic instabilities. LES is utilized to thoroughly analyze the Continuously Variable Resonant Combustor (CVRC)[16], a self-excited unstable test case from Purdue University. The literature reports several numerical studies[11,12,14,17–21] that examine CVRC instability. The pertinent physics underlying the onset of instability in CVRC at specific geometric conditions has been explained. It is hypothesized that features linked to hydrodynamics/vortex shedding lock-on, evolution of the triple flame structure, and ignition delay are responsible for the thermo-acoustic instability in CVRC. The possibility that the matching of injector and chamber frequencies is the primary cause of instability in CVRC has not been discussed or investigated until now. On the other hand, a multi-element rocket scale combustor is used for a detailed analysis of the injector-chamber frequency coupling phenomenon. The German Aerospace Center (DLR) created a multi-element combustor, BKD[7,22,23], which has been investigated numerically in reference[7,24,25]. These analyses identified self-excited combustion instabilities resulting from the interaction between the injector-chamber's longitudinal and transverse frequency coupling. Using an experimental multi-element setup at Purdue University, Philo et al.[26] investigated injector-coupled instabilities. Zhan et al.[27] also studied injector coupled instability in a 10-injector combustor using a flamelet combustion model. In contrast to a single-step chemistry model, their results showed larger amplitude predictions in the flamelet solution. It briefly discusses resonance occurring in injectors due to the longitudinal mode oscillations of the combustor. In addition, Guo et al.[28,29] used a flamelet technique in a hybrid RANS/LES framework to assess thermo-acoustics in a gaseous methane-oxygen single element and multi-element combustor. The frequency matching of longitudinal mode oscillations in the GOx post with transverse pressure oscillations in the combustion chamber was found to be the self-sustained instability mechanism. Injector-chamber coupled high-frequency events have recently been noted in tests of sub-scale thrust chambers driven by LOx-methane and LOx-hydrogen[6,22,23,30]. The main driving mechanism is injector-chamber coupling, the interaction between the combustion chamber's and the LOx injectors' acoustic resonance patterns. High amplitude instability has been found to occur when the chamber's first tangential (1T) mode frequency and the LOx-post second longitudinal (2L) mode overlap. This complex three-dimensional acoustic interaction between the injector and combustion chamber acoustics brings a positive feedback loop, a source of self-sustained instability, and localized variations in heat release. Such complex dynamic properties in a multi-element chamber, which are very difficult to analyze experimentally, can be studied using high-fidelity LES and techniques like dynamic mode decomposition (DMD). The primary goal of this work is to investigate the well-known CVRC problem and to establish a numerical approach for simulating and comprehending the origin of self-excited instabilities. The ultimate goal is to examine the intricate combustion dynamics in a multi-element rocket combustor operating near-ignition conditions, a state characteristic of a full-scale rocket engine utilizing GOx-methane propellant. The objective is to investigate complex processes related to flame-wall contact, multi-element flame interaction, and injector-chamber acoustics. We discuss the dynamic characteristics of injectors, which are sensitive elements and can induce flow and combustion oscillations. The main area of emphasis is to explain the influence of injector-chamber frequency coupling on enhanced combustion dynamics. This research aims to develop a systematic approach and technique for modeling combustion instability and employ data decomposition techniques to examine intricate datasets. This work uses combined LES and DMD analysis to reveal the fundamental interactions between acoustic waves and unsteady combustion heat release. The DMD approach is used to interpret the mode shapes in the combustor and injectors to investigate the possible influence of injector-chamber frequency coupling leading to high-frequency instability in the multi-element combustor. The remaining portion of the paper is structured as follows: first, a concise synopsis of the numerical framework is given; next, the CVRC test case is thoroughly analyzed; and lastly, the identification of the combustion instability mechanism in a multi-element combustor is addressed.

## II. NUMERICAL FRAMEWORK

This section provides a concise overview of the numerical framework utilized in this study. Our earlier study[8] provides detailed information about the numerical framework for combustion dynamics; therefore, the fundamental governing equations are not presented here.

### A. Flow modeling using LES:

LES models the turbulent flow field so that the energetic larger-scale motions are resolved, and only the small-scale fluctuations are modeled. Therefore, the equations solved are the filtered governing equations for conserving mass, momentum, energy, and scalar transport[31]. The subgrid stress modeling is done using a dynamic version of the Smagorinsky-Lilly model by Germano et al.[31] where the unresolved stresses are related to the resolved velocity field through a gradient approximation:

$$\widetilde{u_i u_j} - \tilde{u}_i \tilde{u}_j = -2\,\nu_t\,\bar{S}_{ij} \tag{1}$$

$$\nu_t = C_s^{\,2} (\Delta)^2 \left|\bar{S}\right| \quad \text{, where} \tag{2}$$

$$\bar{S}_{ij} \equiv \frac{1}{2}\left(\frac{\partial \bar{u}_i}{\partial x_j} + \frac{\partial \bar{u}_j}{\partial x_i}\right) \tag{3}$$

$$|\bar{S}| = \sqrt{2\bar{S}_{ij}\bar{S}_{ij}} \tag{4}$$

and S is the mean rate of strain. The coefficient $C_s$ is evaluated dynamically based on the information provided by the resolved scales of motion.

### B. Combustion Modelling:

In this study, fuel and oxidizer enter the reaction zone in separate non-premixed streams, and a conserved scalar mixture fraction approach[32] is used for reactive flow modeling. Our previous work[8,33] highlights the utility of the flamelet approach when turbulent time and spatial scales are bigger than chemical scales. The flamelet-generated manifold approach incorporates realistic chemical kinetic and non-equilibrium effects into the methane-oxygen flame. In this method, the chemistry is pre-processed by solving the diffusion flamelet equations, and a look-up table is generated for mean quantities related to the conserved scalar mixture fraction. Two transport equations for the mean mixture fraction and its variance are solved along with basic governing equations representing $CH_4$-$O_2$ combustion. The details of the combustion model can be accessed from our earlier publication[9]. However, Favre averaged transport equation for the mixture fraction is given as:

$$\frac{\partial}{\partial t}\left(\rho\bar{f}\right) + \nabla.\left(\rho\vec{v}\bar{f}\right) = \nabla.\left(\frac{\mu_{eff}}{\sigma_t}\nabla\bar{f}\right) + S_m \tag{5}$$

here, $S_m$ is only due to the mass transfer into the gas phase from liquid droplets. $\mu_{\mathrm{eff}}$ is the effective viscosity composed of a laminar ($\mu_l$) and a turbulent contribution ($\mu_t$). In this framework, the transport equation for mixture fraction variance is not solved; instead, it is modeled by an algebraic equation given as:

$$\overline{f'^2} = C_{var} L_s^2 \left|\nabla\bar{f}\right|^2 \tag{6}$$

The algebraic form used in this study is based on equalizing production and dissipation at the sub-grid level. This assumption holds true for a refined mesh used in LES. We have observed that grid resolution and time-step size are dominant factors controlling the overall accuracy of the solution rather than the transported or algebraic form of variance formulation. The constant $C_{var}$ is computed dynamically based on the procedure of Germano[31] for the dynamic SGS model. A validated $CH_4$-$O_2$ Chemkin mechanism generates non-adiabatic flamelets, as provided in our earlier work[33]. All governing equations are implicitly filtered by the finite-volume methodology of ANSYS Fluent[31]. Spatial discretization is performed by second-order bounded schemes, with time integration using the bounded second-order implicit method.

## III. DYNAMIC MODE DECOMPOSITION

Dynamic Mode Decomposition (DMD) is a powerful technique to extract coherent structures and dynamic patterns from high-dimensional, time-varying data. It is a data-driven approach beneficial for analyzing the spatio-temporal evolution of fluid flow fields. This method involves analyzing temporal data to extract coherent structures and dominant modes, facilitating understanding of system behavior and dynamics. Comprehensive insights into this technique can be found in Schmid[34] and Rowley et al.[35]. DMD is advantageous, providing a compact representation of complex fluid flow data and identifying dominant patterns and structures. The Arnoldi algorithm reduces the data in DMD, generating decomposed modes based on discrete frequencies. DMD calculates the temporal transition between snapshots to identify coherent structures corresponding to one frequency. The advantage of DMD lies in its organization of DMD modes based on individual frequency, in contrast to POD, where the distinct impact of an individual frequency is not identifiable. It enables the visualization of distinct flow field structures corresponding to individual frequency modes. However, POD has been effectively applied to specific cases highlighted in references[36–38]. The eigenvalues within DMD signify the growth, decay, and oscillation frequencies inherent in each mode. Each decomposed mode can be represented in terms of a spatial and a temporal response, which provides detailed insight into the dynamics and offers a systematic way to explore crucial physics within coupled acoustics and combustion scenarios. Huang et al.[39] extensively compared POD and DMD methodologies to analyze self-excited combustion instability. Their findings highlighted the suitability of DMD for examining periodic combustion instability. They underscored that DMD is a more systematic and efficient tool for extracting and isolating specific phenomena from comprehensive datasets containing multiple physical occurrences. The DMD method has been widely utilized for identifying the spatial modes of hydrodynamic and thermo-acoustic instability[8,28,29,40]. Recent developments have introduced various forms of DMD; however, the method proposed by Sayadi and Schmid[41] is used in this study. Bhatia et al.[42,43] effectively employed this method to study the complex dynamics of spray flame combustion. In the DMD method, N number of data sets or snapshots equally separated in time can be arranged in a data matrix $V_1^N$, whose columns are the individual snapshots/data sets.

$$V_1^N = \{v_1, v_2, v_3 \dots v_N\} \tag{7}$$

These snapshots/data sets are assumed to be related via. linear mapping, which defines a linear dynamical system. DMD assumes that a linear map exists between a snapshot and the following snapshot in the sequence, such that:

$$v_{i+1} = Av_i \tag{8}$$

where it is assumed that $A$ is a matrix that linearly maps the $i^{th}$ sequence to the $(i + 1)^{th}$ field. In matrix form, $V_1^N$ can be represented as:

$$V_1^N = \{v_1, Av_1, A^2v_1 \ldots A^{N-1}v_1\} \tag{9}$$

The eigenvalues and vectors of matrix $A$ can describe the dynamic characteristics of the data sequence. Another assumption exists in which the vector $v_N$ can be expressed as a linear combination of the previous vectors till a specific number N. The vector $v_N$ can be represented as a linear combination of previous vectors as:

$$v_N = a_1v_1 + a_2v_2 + \cdots + a_{N-1}v_{N-1} + r = V_1^{N-1}a + r \tag{10}$$

$a = a_1, a_2, a_3 \ldots, a_{N-1}$ represents the coefficients of DMD, and $r$ is the residual vector. The above equation can be recast as:

$$V_2^N = AV_1^{N-1} = V_1^{N-1}S + r.e_{N-1}^T \tag{11}$$

where, $V_2^N = \{v_2, v_3 \ldots v_N\}$

In this context, S represents an unknown companion matrix where $a$ is determined by minimizing the overall residual $r$ via the solution of a least squares problem using QR factorization. When an ample number of snapshots are utilized, the eigenvalues of $S$ effectively represent the eigenvalues of $A$, capturing the time-evolution characteristics of the flow field. Consequently, the eigenvalues and eigenvectors of $S$ define the dynamic modes inherent to the considered flow field. The $k^{th}$ dynamic mode, $\varphi_k$ corresponding to the frequency response can be constructed as:

$$\varphi_k = AV_1^{N-1}y_k \tag{12}$$

Where, $y_k$ is the $k^{th}$ eigenvector of matrix S. The original data set can be decomposed into

$$V_1^{N-1} = \sum_k \varphi_k y_k^T \tag{13}$$

Where, $\varphi_k$ contains the dynamic spatial information, and $y_k$ contains the temporal evolutional information. More details on DMD applications can be found in the earlier works of our group[36,40,42–45]. This study aims to use the DMD to reveal the physics underlying the combustion dynamics in a rocket scale combustor. The main focus of DMD analysis is to determine the mode structure in the chamber and injector regions, aiming to identify potential frequency coupling between the injector and chamber. This coupling could significantly contribute to enhanced combustion dynamics in the cases studied.

## IV. CONTINUOUSLY VARIABLE RESONANT COMBUSTOR (CVRC) TEST CASE

Before explaining simulation results, we briefly overview the CVRC test case. A third-generation single-element rocket combustor known as a continuously variable resonance combustor was created by Yu et al.[16]. Its purpose was to create and maintain longitudinal mode instabilities in a sub-scale rocket combustor. This experimental work is an excellent starting point for developing and validating a numerical approach to model and understand thermo-acoustic instability in a typical rocket chamber. Several research teams have used high-fidelity simulations to study combustion dynamics in CVRC. A summary of the numerical studies conducted on CVRC is provided here for completeness. Garby et al.[17] used the AVBP LES algorithm to study CVRC. It used a two-step reaction mechanism with a dynamically thickened flame (DTF) model. This approach recovers the sub-grid reaction rate using an efficiency function that artificially thickens the flame to match the LES mesh resolution. The evolution of the triple-flame structure supported by burned gas and flame stabilization was associated with the self-excited instability process. Srinivasan et al.[14] used the LESLE algorithm with the linear eddy model (LEM) for turbulent combustion to simulate the CVRC under stable and unstable operating conditions. The pressure oscillation amplitude was underestimated, and the dominant frequency was overestimated. Using a single-step chemistry, Harvazinski et al.[18] conducted 2D and 3D simulations using the GEMS[46] code. The mechanism of self-excited instability in CVRC is investigated using a hybrid detached eddy simulation (DES) technique. By analyzing the relationship between hydrodynamics, acoustics, and heat release, he gave a thorough understanding of the causes of combustion instability. Three main mechanisms of instability were found in the study: the presence of a triple-flame structure, improved mixing caused by baroclinic torque, and pulse timing. The unsteady pressure cycle analysis clarified the identified instability mechanisms. A kinetic model and the flamelet approach were thoroughly compared in a recent work by Harvazinski et al.[19]. The results show that flamelet and finite rate computations produce comparable results. The need to use a detailed chemistry mechanism to accurately predict heat release, ignition delay, and subsequently self-excited instability in CVRC was highlighted by Sardeshmukh et al.[21]. Compared to 2D simulations utilizing a single-step reaction, the detailed chemistry axisymmetric DES was able to capture oscillation amplitude accurately. Using delayed detached eddy simulation, Nguyen et al.[20] investigated the CVRC with a flamelet-progress variable combustion model. The axisymmetric simulations demonstrated reasonable agreement in frequency but underestimated the amplitude. A 2D modeling framework was developed by Pant et al.[12] integrated with hybrid RANS/LES. The study emphasized the significance of ignition delay and transient flame dynamics as possible causes of both self-excited and sustained thermo-acoustic instability in CVRC. Although various studies have been conducted on this benchmark problem, relatively few investigate the possibility of injector-chamber frequency coupling as the cause of instability in CVRC. This aspect of high-frequency instability has not received much attention in numerical studies despite significant experimentation conducted at DLR[47]. There is a clear need to explore and understand this aspect of high-frequency instability. To ascertain whether injector-chamber frequency coupling could result in self-sustaining oscillations, we examine the CVRC test case in this section. It also serves as a validation for the multi-element combustor presented as another case study. LES is performed on a highly unstable CVRC test scenario, utilizing the numerical framework developed in our earlier work[8]. The flamelet-based methodology is applied to capture combustion dynamics, which is computationally less demanding and integrates the effects of turbulence-chemistry interaction with detailed finite-rate chemical kinetic effects. Further details on this methodology can be found in references[8,13,48].

### A. Physical model and boundary conditions

Most numerical studies related to CVRC usually start with 2D modeling and move on to 3D simulations using simplified oxidizer and fuel inlet boundaries. However, to properly replicate the intricate geometries of the oxidizer and fuel inlet, a three-dimensional model of the combustor, including an intermediate oxidizer post length of 12 cm, is created for this study. This approach mimics the impedance boundary conditions observed in the tests. This study involves the simulation of a highly unstable benchmark case introduced by Yu et al.[16] to replicate the fluctuating pressure behavior observed within the specific operational parameters of CVRC. A translating shaft is used to vary the oxidizer-post length (Lop) in the CVRC experiments. This range of variation spans from Lop = 19.05 cm to Lop = 8.89 cm and vice versa, allowing the combustor to operate in stable and unstable environments. The combustor exhibits limit cycle oscillation in unstable conditions, frequently achieving peak-to-peak pressure amplitudes up to 50% of the mean chamber pressure. There is a discernible increase in the magnitude of instability in the range of oxidizer post lengths from 10 to 16 cm, with the maximum amplitude at 12 cm. Further details on the test setup and test results can be found in Yu et al.[16]. Most computational efforts have focused on simulating the conditions with fixed oxidizer post lengths of 12 or 14 cm. The oxidizer post length of 12 cm shows the maximum power spectral density (PSD) and is consistently unstable in all tests with no noticeable hysteresis effect. In this case study, the LES framework of ANSYS Fluent 2023R1[49] is used to perform a thermo-acoustic simulation of the CVRC for an oxidizer post length of 12 cm. The combustor's geometric details, such as the injector shape, fuel inlet manifold, and oxidizer slotted plunger, are shown in Figure 1. An enlarged picture of the oxidizer's slotted plunger reveals that the fuel enters the domain through 36 holes. A coaxial injector, with the fuel at the periphery and the oxidizer in the center, introduces reactants into the chamber. Methane serves as the fuel, while catalyst-decomposed liquid hydrogen peroxide produces hot steam and oxygen, used as the oxidizer. During testing, a slotted inlet design is employed for the oxidizer entry to maintain the choked inlet condition. The computational domain is truncated at the nozzle throat, ensuring choked/full reflective outlet conditions during simulation.

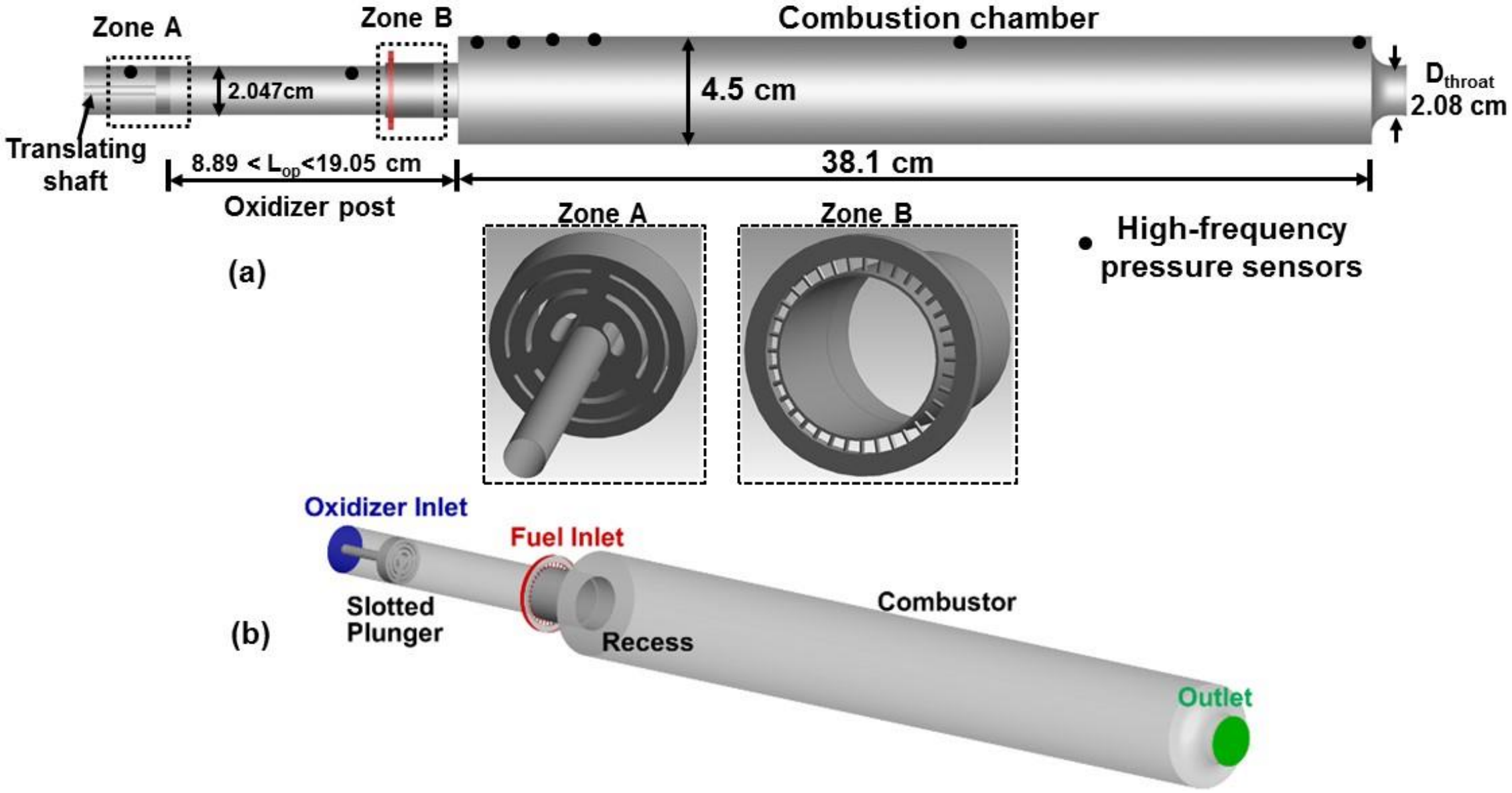

**FIG.1.:** Close up view of oxidizer plunger (a) & fuel Inlet Section (b) Computational domain CVRC

In CVRC LES, mass flow inlet boundary condition is imposed for $CH_4$ with a total mass flow rate of 0.0272 kg/s, injected at a temperature of 300K, whereas oxidizer comprises 42% $O_2$ and 58% $H_2O$, at a mass flow rate of 0.3180 kg/s and injection temperature of 1030 K. Combustor and injector walls are treated as adiabatic. The global equivalence ratio for the test simulated is $\Phi$ = 0.813. Before commencing the LES case, a RANS simulation is conducted to verify the choked condition at the oxidizer plunger and the outlet throat section. The steady-state simulation confirms that the upstream pressure of the oxidizer evolves to 42 bar, closely aligning with the experimental value of 42.93 bar. This observation indicates the presence of a choked oxidizer plunger and chamber throat, ensuring fully reflective acoustic boundary conditions during the execution of LES. We tested three grid sizes, with mesh counts of 6, 7, and 8 million, respectively, intending to capture vortex formation and the propagation of acoustic waves within the chamber. The grid employing a mesh count of 7 million successfully resolves over 80% of turbulent scales and is chosen for further analysis in this study. Figure 2 illustrates the resolved kinetic energy spectrum displaying a theoretical slope of -5/3 in the inertial subrange and steep decline at higher frequency. This outcome illustrates the grid's capability to distinguish between integral and Kolmogorov length scales, confirming the adequate resolution of scales in the conducted LES.

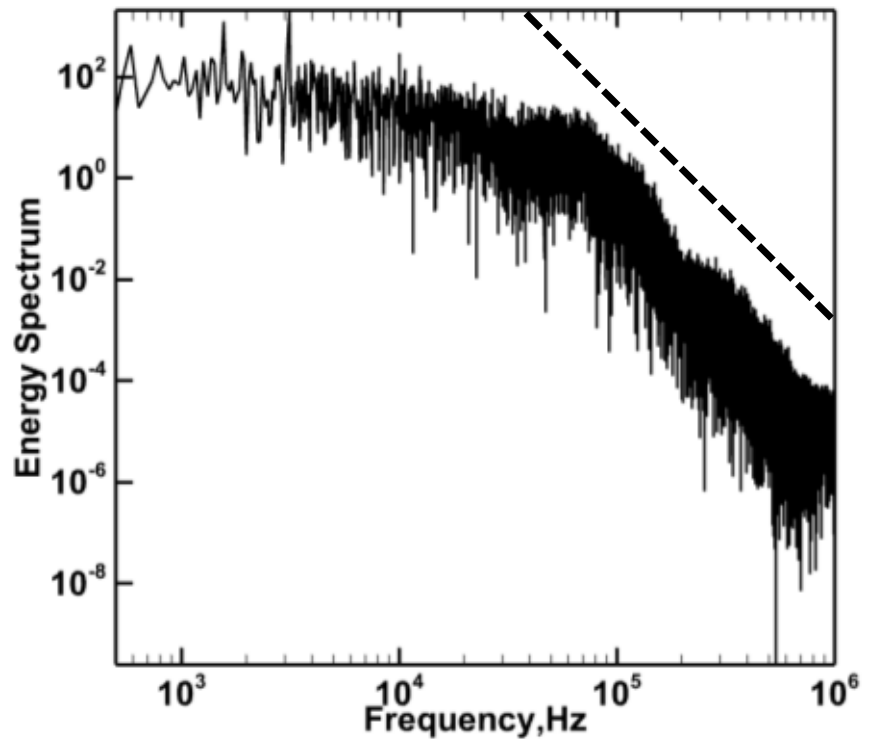

**FIG.2.**: Resolved kinetic energy spectrum

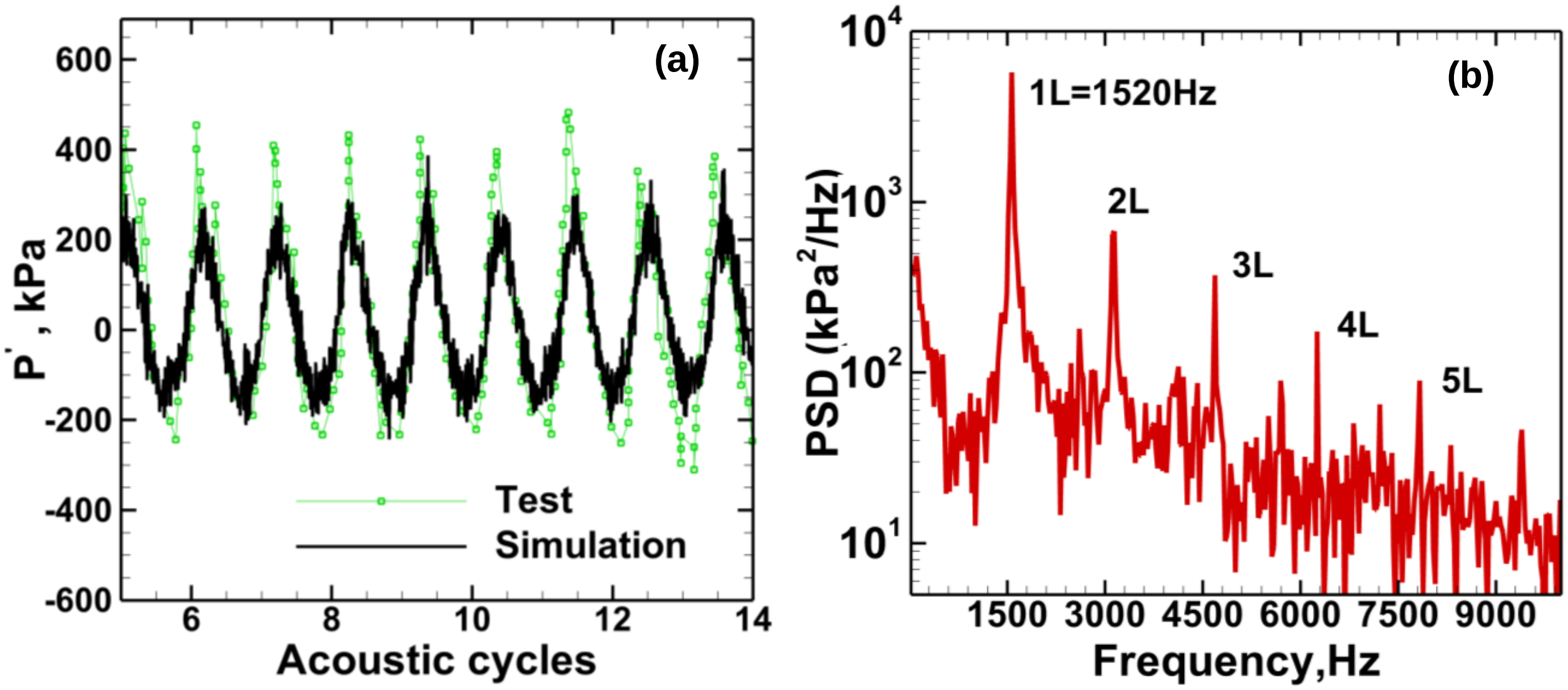

**FIG.3.**: (a) Peak to peak raw pressure fluctuations. Test results are measurements[14]
(b) Harmonics of raw pressure data

## C. Instability characteristics in CVRC

We conduct LES for over 50 acoustic cycles, ensuring the overall acoustic CFL (ACFL) is below 4. A time step size of 0.5μs is used in this study to capture the acoustic activity. We initially present the comparison of peak-to-peak amplitude of raw pressure fluctuations from LES with experimental data[14]. The comparison of the test and simulation pressure variation at the combustor end point is shown in Figure 3(a). The LES pressure variation displays the self-excited limit cycle behavior, closely matching the test data. We have noticed a peak-to-peak pressure fluctuation amplitude of 5 bar in the simulation, which amounts to 30% of the mean chamber pressure. A reasonable match is observed between the test and simulation, with an underprediction of pressure amplitude on the positive side of the spectrum. Figure 3(b) displays a spectral plot of pressure data, which depicts the harmonics of the first longitudinal mode in the combustor. The spectral plot shows a well-organized pressure oscillation in the combustor. LES displays a 1L mode at 1520Hz, with second harmonics (2L) present near 3100Hz and so on, while the test exhibits a 1L mode close to 1400Hz [16] .

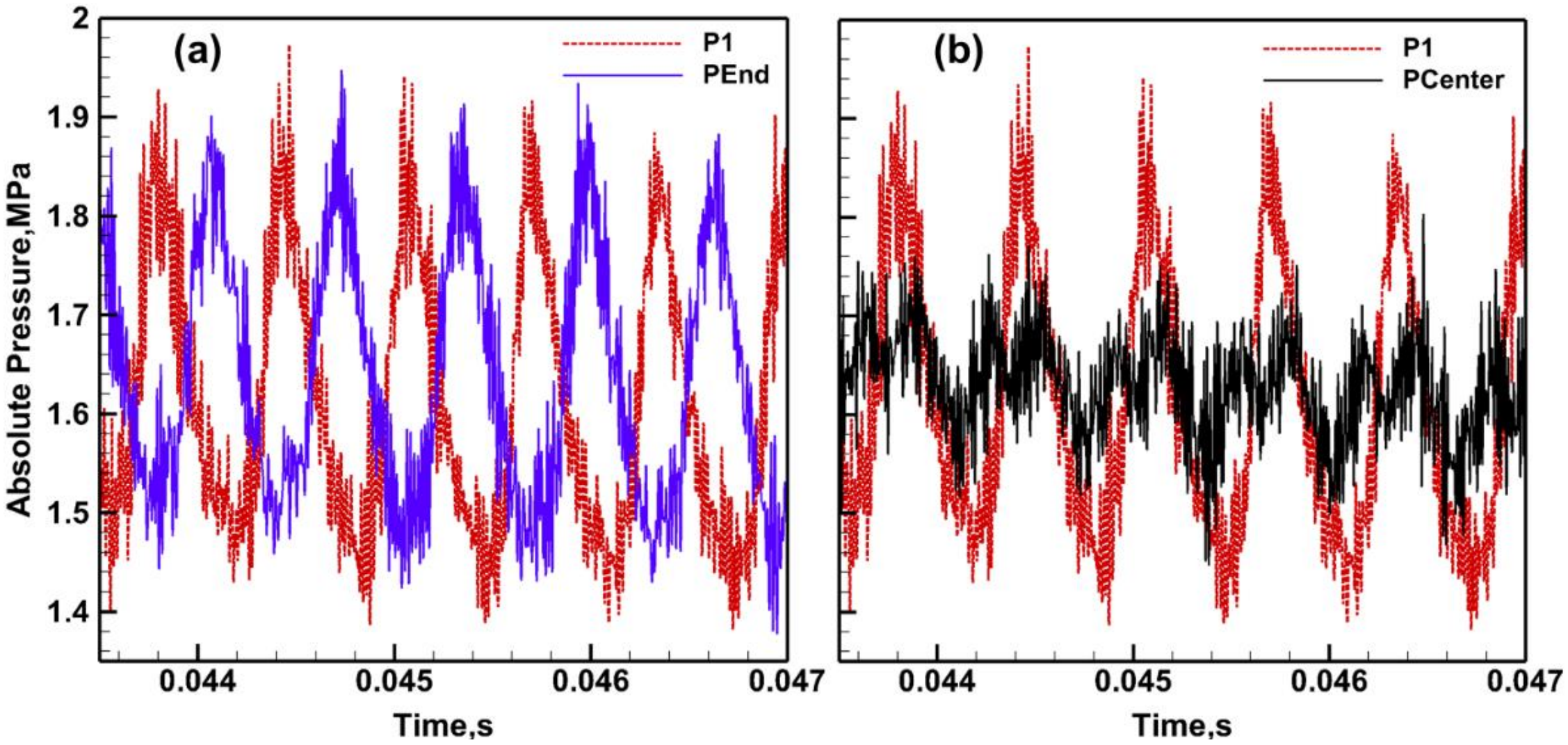

**FIG.4.**: (a) Pressure comparison P1-Pend and (b) P1-Pcenter

Due to the simulations neglecting heat loss and assuming 100% combustion efficiency, the 1L mode is overestimated. The elevated operating pressure and higher burnt product temperature lead to increased sound speed, ultimately resulting in an overestimation of the eigenmode frequency. Studies on the CVRC combustor in the literature have also shown similar overestimations of frequency caused by these factors [14,17]. The LES methodology effectively captures self-sustained instability, as evidenced by the acceptable agreement between LES and test data in pressure fluctuation amplitude and dominant frequency. Additionally, this scenario provides an excellent opportunity to determine the mechanisms causing instability in CVRC. To understand the movement of the acoustic waves in CVRC, we examine the pressure probe data. The pressure variation at the P1 (chamber start) and Pend (chamber end) probe locations are shown in Figure 4(a). The out-of-phase pressure variation indicates that the high-pressure zone moves from the head end downward. The out-of-phase pressure variation also displays the longitudinal nature of the acoustic wave in CVRC. Figure 4(b) further illustrates the variation between P1 and the probe at the centre of the combustor (PCenter). It displays the anti-nodal and nodal pressure characteristics at the respective probe locations, displaying a typical nature of longitudinal wave movement.

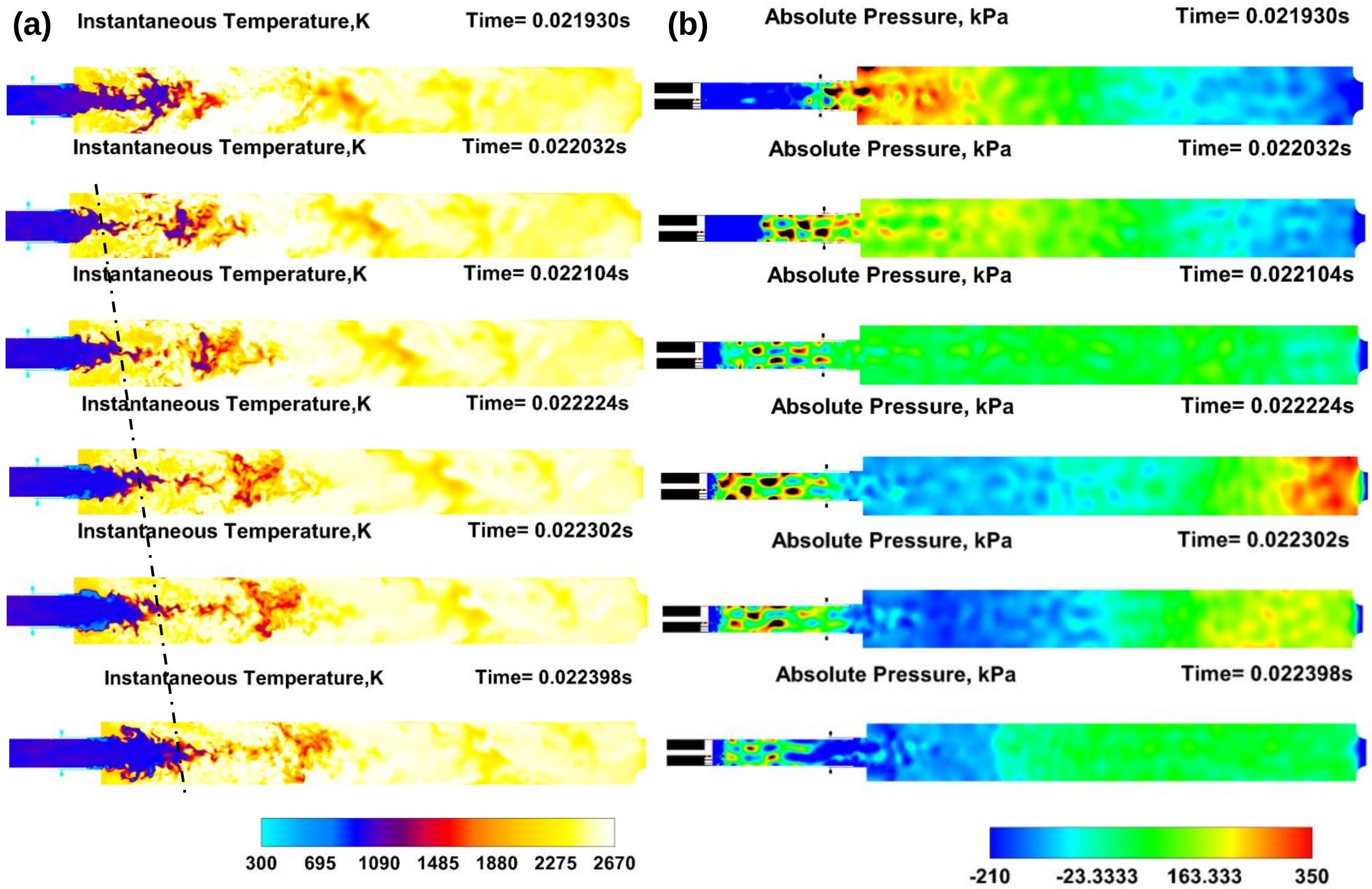

**FIG.5.**: Instantaneous variation with acoustic wave movement towards down-end: (a) Temperature, (b) Pressure

Here, we briefly discuss the impact of longitudinal acoustic waves on the flame features. The relevant flow conditions that lead to self-sustained acoustic waves within the combustor are discussed. The instability mechanism is further explored and discussed based on the analysis of existing theories and exploration of injector-chamber frequency coupling. Figure 5 displays instantaneous flame temperature variation at different time instants, with pressure wave movement from head to down end region. Instantaneous temperature contours in Figure 5(a) display the flame structure with the movement of a pressure wave away from the head end. Figure 5(b) depicts the pressure movement in the combustor. It illustrates the flame propagating downstream as the pressure wave moves away from the head end region. It illustrates the effect of acoustic wave movement within the combustor on the flame characteristics. An axial stretch of the flame is noticed, with pressure wave movement towards the down end segment of the combustor, with a dotted line depicting the axial stretch of the flame in Figure 5(a). Figure 6 illustrates the instantaneous variation in methane and oxygen concentrations in the combustor as the pressure wave moves towards the down end. Figure 6(a) shows a significant increase in methane concentration as the pressure wave moves away from the head region. A higher mass fraction during low pressure at the head end can lead to fuel accumulation in the back step region. This accumulated fuel can violently burn during the return of the pressure wave, encouraging higher pressure oscillations. A similar variation in oxygen concentration is also observed, as depicted in Figure 6(b).

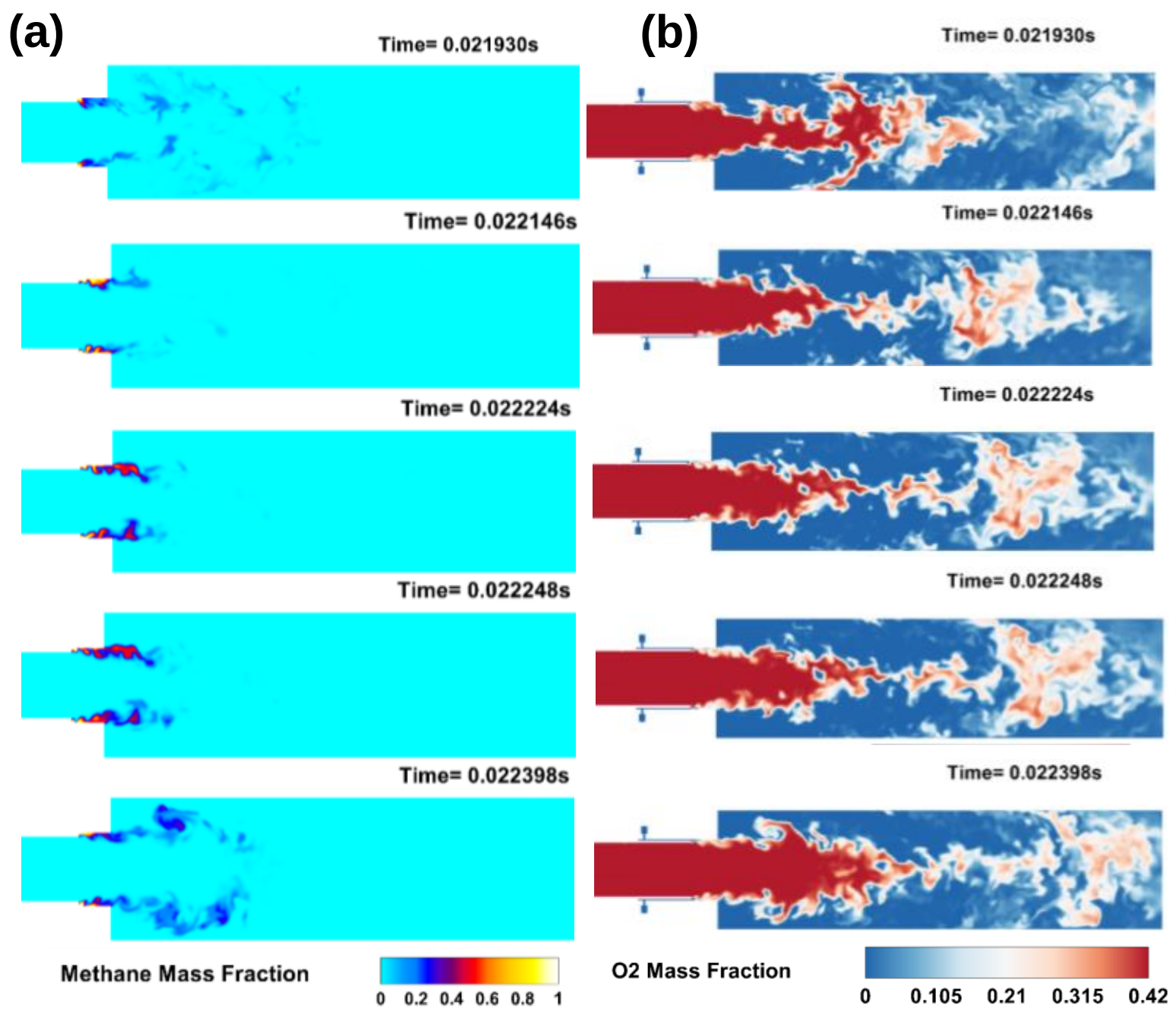

**FIG.6.**: Instantaneous concentration variation with acoustic wave movement towards down-end: (a) methane, (b) oxygen

Our previous study[13] covered the instantaneous flow features of the unstable scenario and depicted the flow characteristics of methane and oxygen entering through 36 orifices and slotted plungers, respectively. At the oxidizer entry, the slotted plunger maintains an acoustically choked condition. In the back step region, a corner recirculation zone (CRZ) forms, which causes the flame to stabilize. Product species are produced at a shear layer resembling the structure of a diffusion flame, and their concentrations increase along the length of the combustor. The concentration of OH is frequently used as a heat release indicator. Figure 7 displays the instantaneous contour of the OH mass fraction and progress variable in one instability cycle at the axial plane of the combustor. The instability cycle starts ($0^{o}$) and ends ($360^{o}$) with high pressure at the head end. Figure 7(a) shows the attachment and detachment of flame through OH concentration.

The OH mass fraction initiates downstream of the injector exit in the shear layer. The OH concentration at 0° and 60° signifies an attached flame stabilized in the back step region. The flame detaches at 180° during a low-pressure condition at the head end and reattaches to the oxygen post at 360° during the return of the high-pressure wave to the head end. The attached and detached nature of the flame can result in heat release rate oscillations.

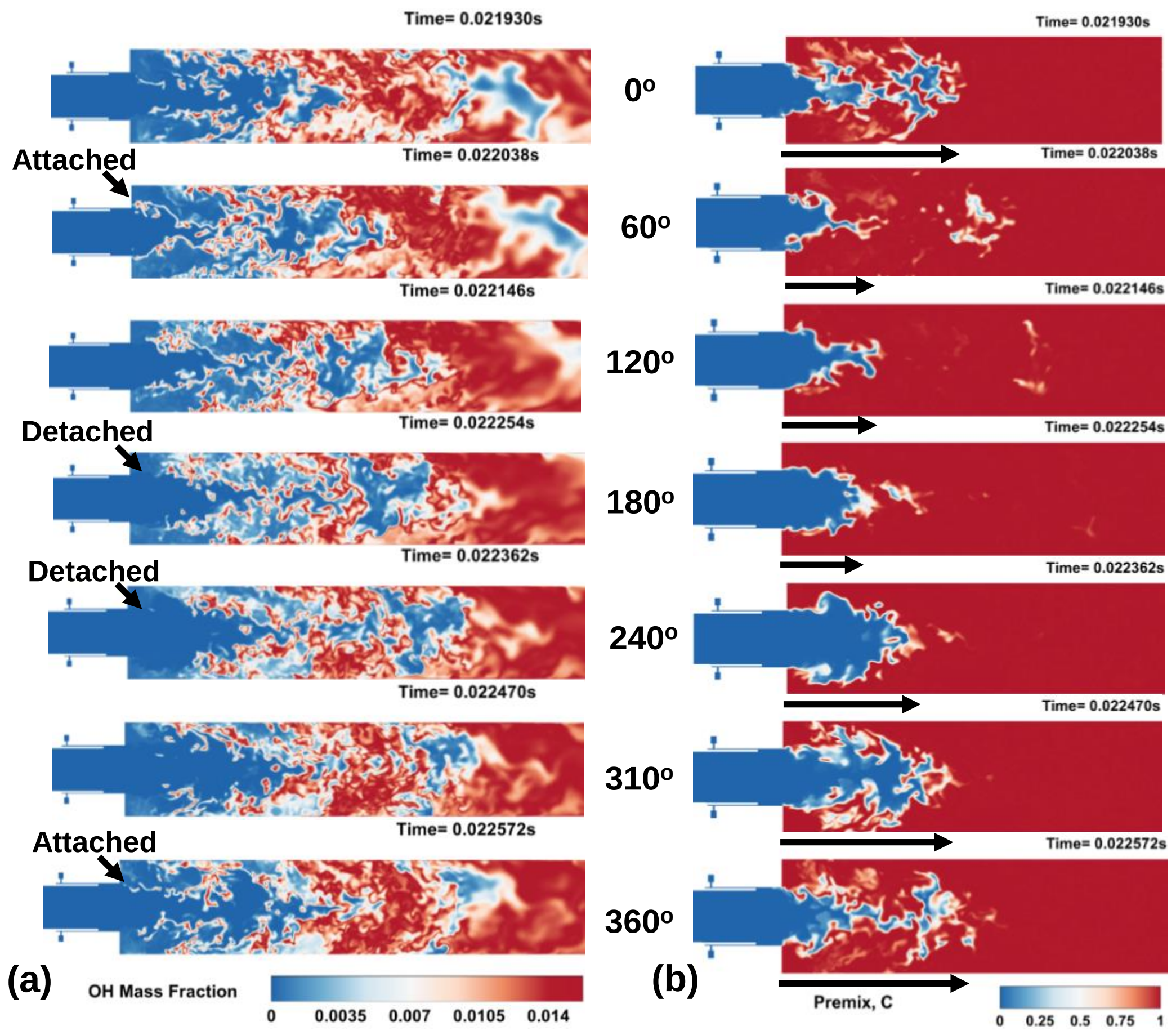

**FIG.7.**: Variation in a single instability cycle: (a) OH (b) Progress variable

Figure 7(b) displays the variation of the progress variable during the instability cycle. It depicts the outward movement of the cold stream (unburnt) during the away movement of the pressure wave, whereas the burnt flame surface area increases during the return of high pressure at the head end. It should be noted that the progress variable formulation in the FGM (Flamelet Generated Manifold) model could capture the flame attachment, detachment, and reattachment at the oxygen post location. This combustor's non-equilibrium (kinetic) effects are well captured through the FGM model, which may not be possible with standard non-premixed flamelet formulation.

## D. Instability mechanisms in CVRC

The mechanism of self-sustained instability is illustrated through the evolution of vorticity, pressure, velocity, and mass flow rate at an injector exit location. The inter-relation of vorticity and pressure wave is presented initially. The aim is to identify the mechanisms responsible for instability in CVRC combustor and scrutinize the already identified mechanisms in the literature. Figure 8 displays vorticity variation at the head end during the presence and absence of a high-pressure wave. It shows a parametric comparison at two different timestamps beginning from the lowest to the highest pressure at the head end. The evolution of vorticity magnitude is shown in Figures 8(a) and 8(b), corresponding to the timestamp of 0.022248s (lowest pressure) and 0.022572s (highest pressure). It exhibits lower vorticity generation at lower backstep pressure and higher vorticity at higher backstep pressure. However, higher methane entry into the chamber occurs during lower back step pressure. The higher vorticity magnitude manifests from the rapid combustion during the higher pressure at the head end. The higher heat release during the high pressure at the head end can lead to higher velocity gradients and, eventually, higher vorticity.

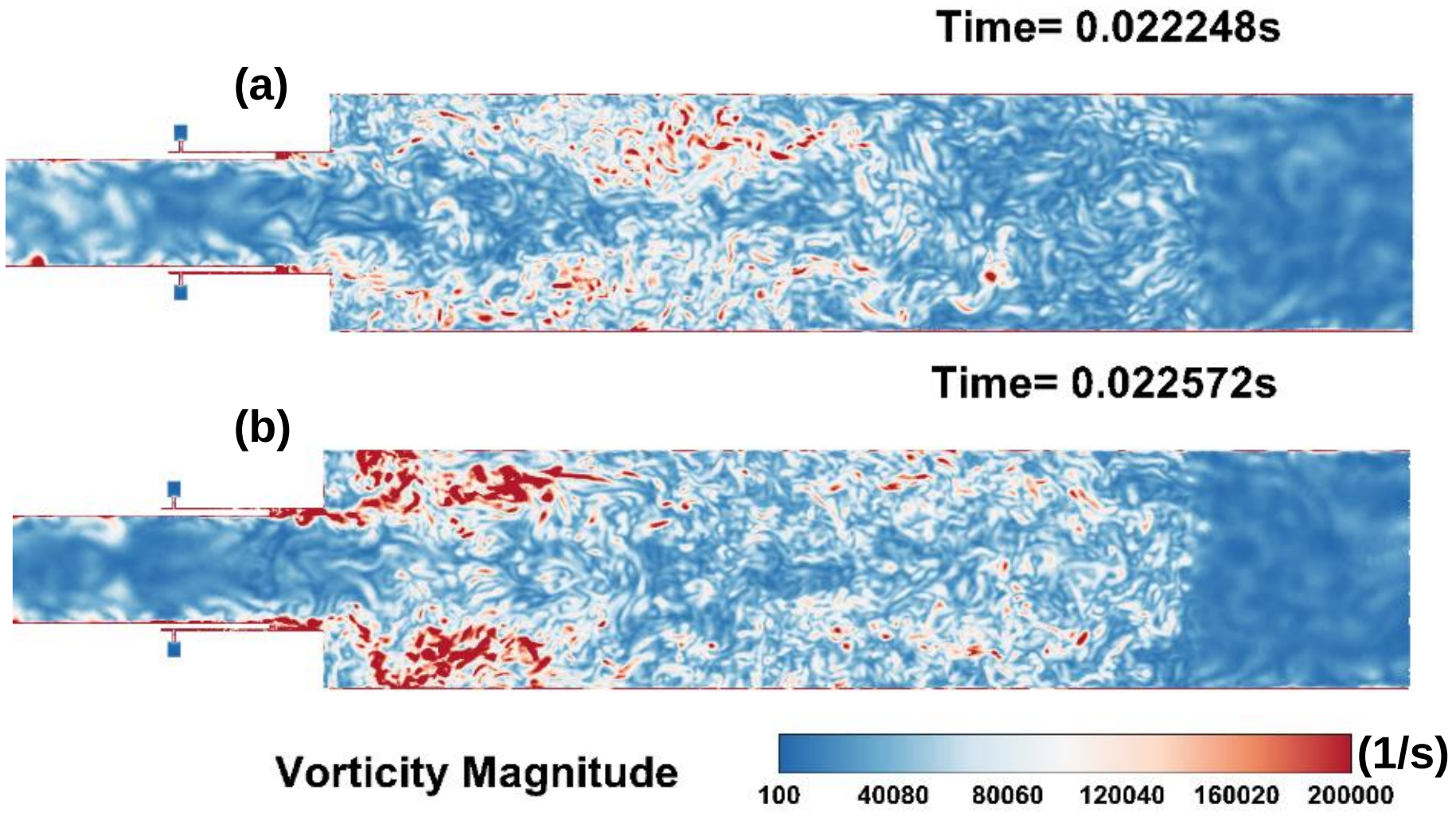

**FIG.8.**: Vorticity variation during high pressure at (a) down end (b) head end

Figure 9(a) displays the variation of instantaneous vorticity and pressure at the head end probe location. It displays pressure and vorticity at a point in the shear layer. The variation shows increased vorticity with increased pressure, attributed to sudden heat release. Figure 9(b) shows the FFT of the pressure and vorticity spectra. It displays the same frequency peak at 1500 Hz, indicating the vorticity-acoustic frequency coupling. It is well known that the rapid changes in pressure, temperature, and heat release induced by vorticity can lead to pressure oscillations and instability. We determine the contribution of hydrodynamic instability to the self-excited oscillations by applying the Strouhal number. It is generally known that acoustic fluctuations can couple with the periodic instability of vortex shedding. This can lead to self-excited oscillations in dump combustor configurations such as CVRC. Here, the Strouhal number is St = fvD/U, where fv is the frequency of vortex shedding, D is the jet flow characteristic length (taken as the diameter of the injector exit), and U is the jet velocity (taken as the facet-averaged axial velocity of the injector exit).

St in the range of the preferred mode close to 0.175 demonstrates intensive mixing under the effect of vortex structures in the shear layer. ~~The vortex shedding frequency at this condition is 1510Hz, which is very close to the first longitudinal acoustic mode of CVRC and indicates formation of vorticity-acoustics feedback loop~~. At this condition, the vortex shedding frequency is 1510Hz, which closely matches the frequency of the first longitudinal acoustic mode of the CVRC, indicating the formation of a vorticity-acoustics feedback loop.

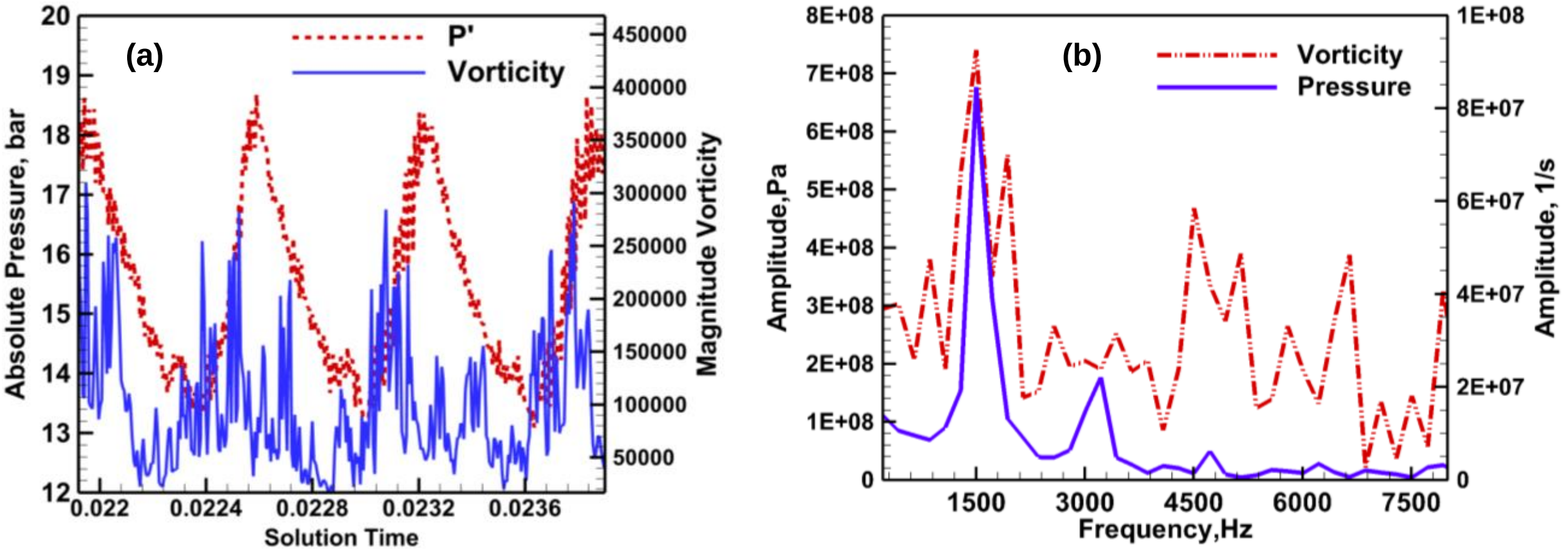

**FIG.9.**: (a) Vorticity variation with pressure, (b) FFT of pressure and vorticity spectrum

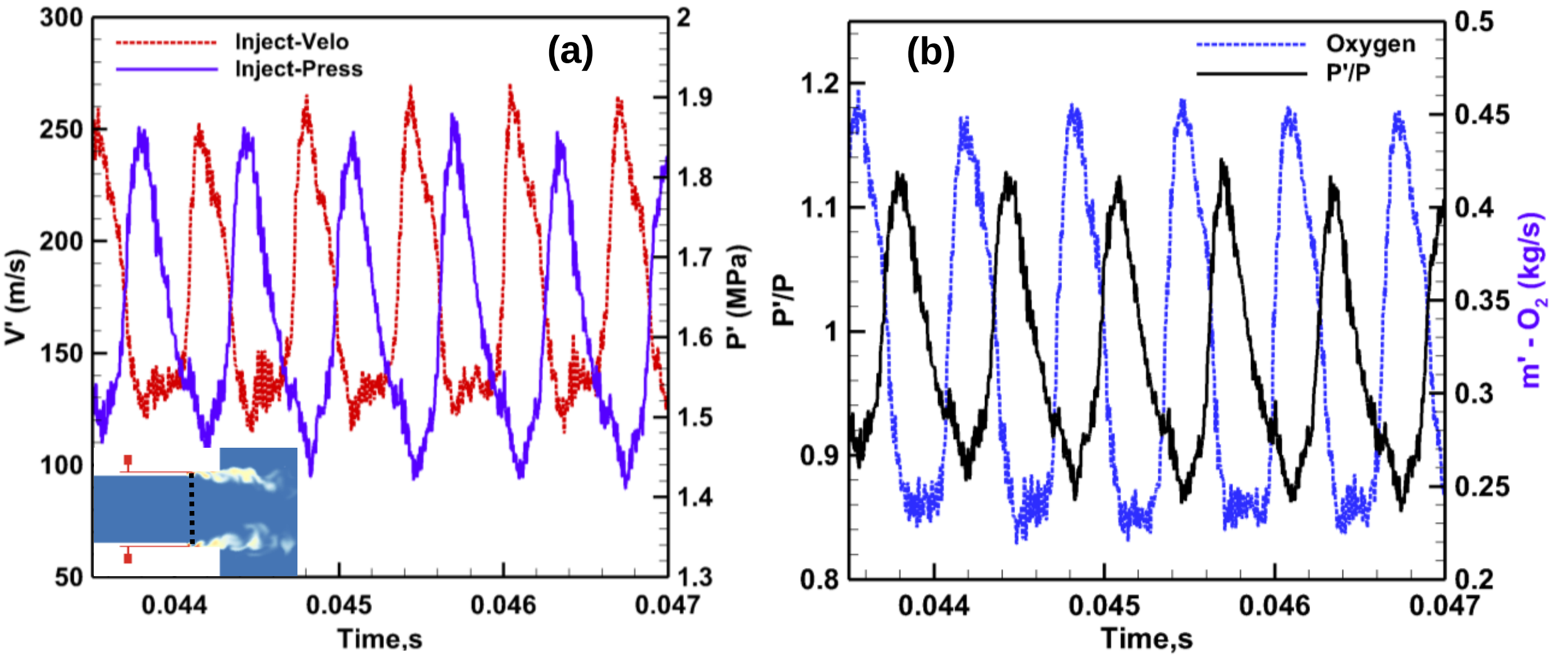

**FIG.10.**: (a) Velocity and pressure variation in the injector, (b) Oxidizer pressure and mass flow rate

The pressure and velocity at the injector can influence the oxidizer and methane mass flow rate, inducing instability, as observed in CVRC. Figure 10 shows the velocity and pressure variation at an injector plane, as depicted in Figure 10(a). Figure 10(a) displays the opposite phase variation for velocity and pressure at the radial plane. The sudden drop in axial velocity indicates that the traveling waves in the oxidizer post periodically prevent the oxidizer from being supplied to the combustion chamber. Figure 10(b) presents the variation of instantaneous oxidizer mass flow rate and normalized pressure at the radial plane. Figure 10(b) displays a sudden oxygen mass flow rate drop during a pressure peak. The sudden drop in oxidizer mass flow rate manifests from the periodic blocking of the oxygen stream by incoming pressure waves towards the head end of the combustor. This effect is further elaborated in methane mass flow rate variation at the injector exit and shear layer location. Figure 11(a) displays methane mass flow rate and pressure variation at the injector exit location. It also shows a similar pattern as the oxidizer mass flow rate. The variation in methane mass concentration in the shear layer illustrates the periodic blocking of methane flow. Figure 11(b) shows methane mass concentration with pressure at probe (Ra) downstream of the injector exit. It shows the sudden drop of methane mass concentration close to zero during the presence of a high-pressure region at the head end. It indicates a fuel cut-off event during the arrival of a pressure wave at the head end. The fuel-cut-off event can instantaneously lead to a very high mixture ratio (mass ratio of oxidizer to fuel), which can trigger and sustain instability in CVRC.

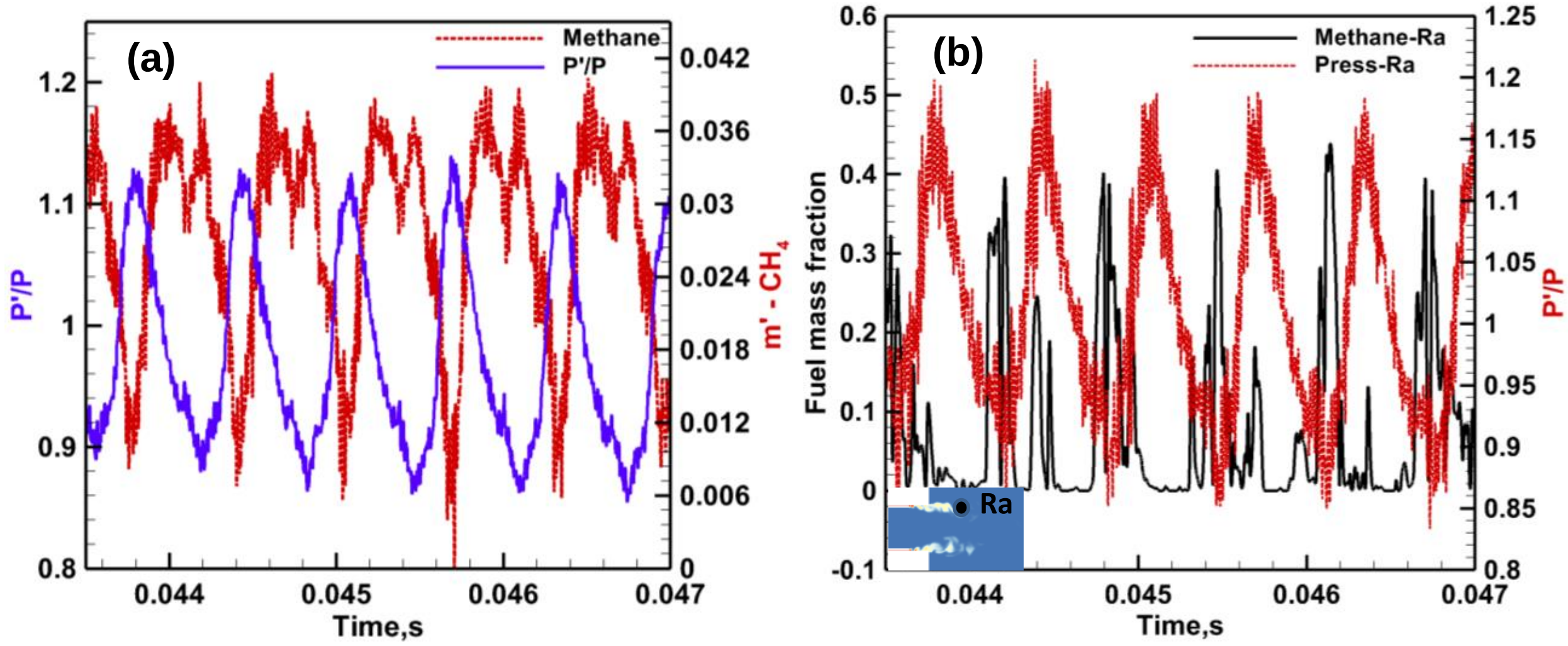

**FIG.11.**: (a) Methane mass flow rate (kg/s) and pressure variation at the injector exit location, (b) Methane mass fraction and pressure variation in the shear layer

Figure 12 shows the comparison of methane and oxygen mass flow rates. It shows that the methane flow rate recovers faster than the oxygen flow rate from a high-pressure wave at the head end. The early recovery can lead to the accumulation of methane in the backstep region, which can ignite suddenly and sustain the thermo-acoustic instability.

Any change in oxidizer post length can change this fuel accumulation characteristic, and CVRC may not show self-sustained oscillations, as seen in this oxidizer tube length. Higher heat release at the pressure antinode location provides the potential to amplify instability.

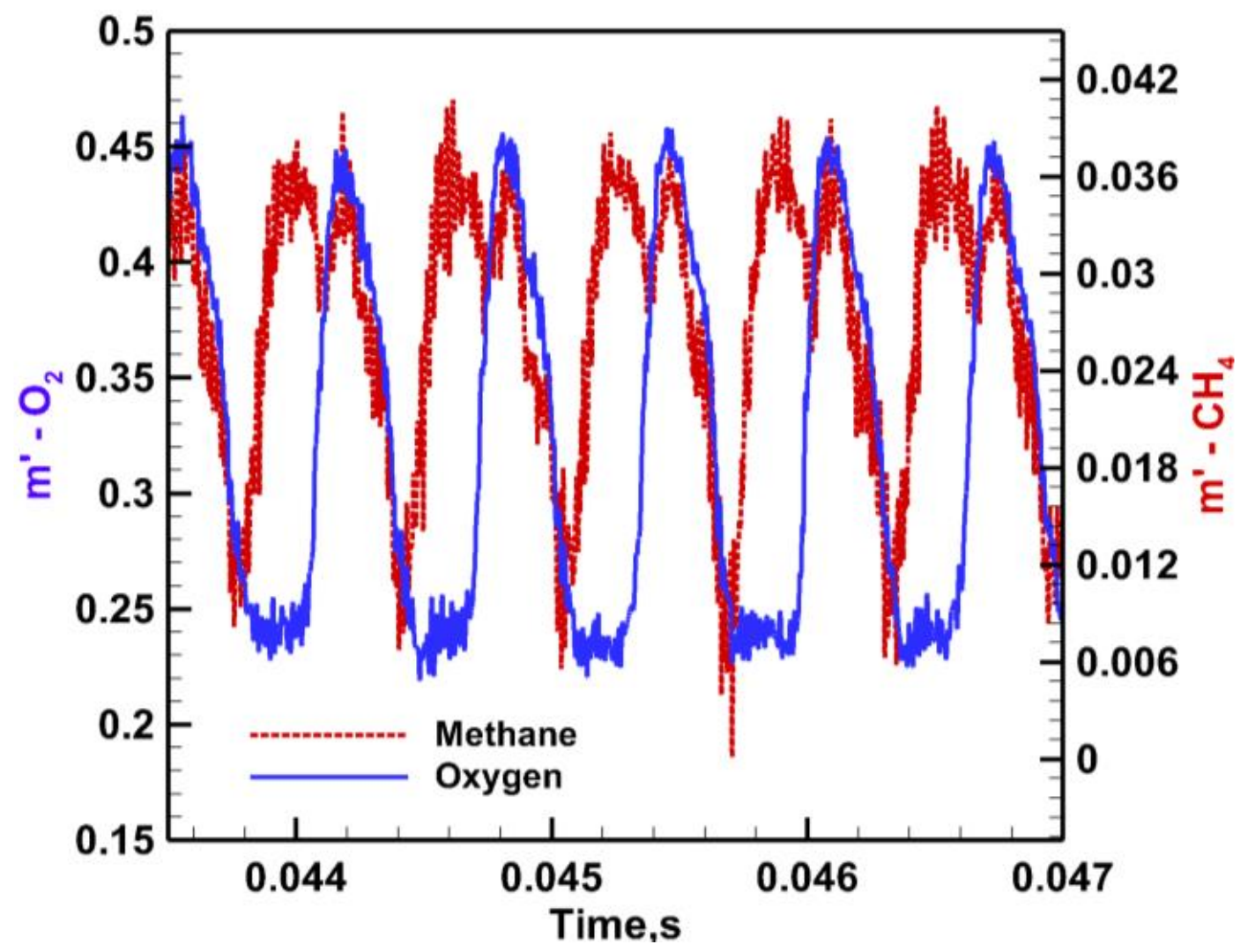

**FIG.12.**: Combined methane and oxygen mass flow rate (kg/s) variation

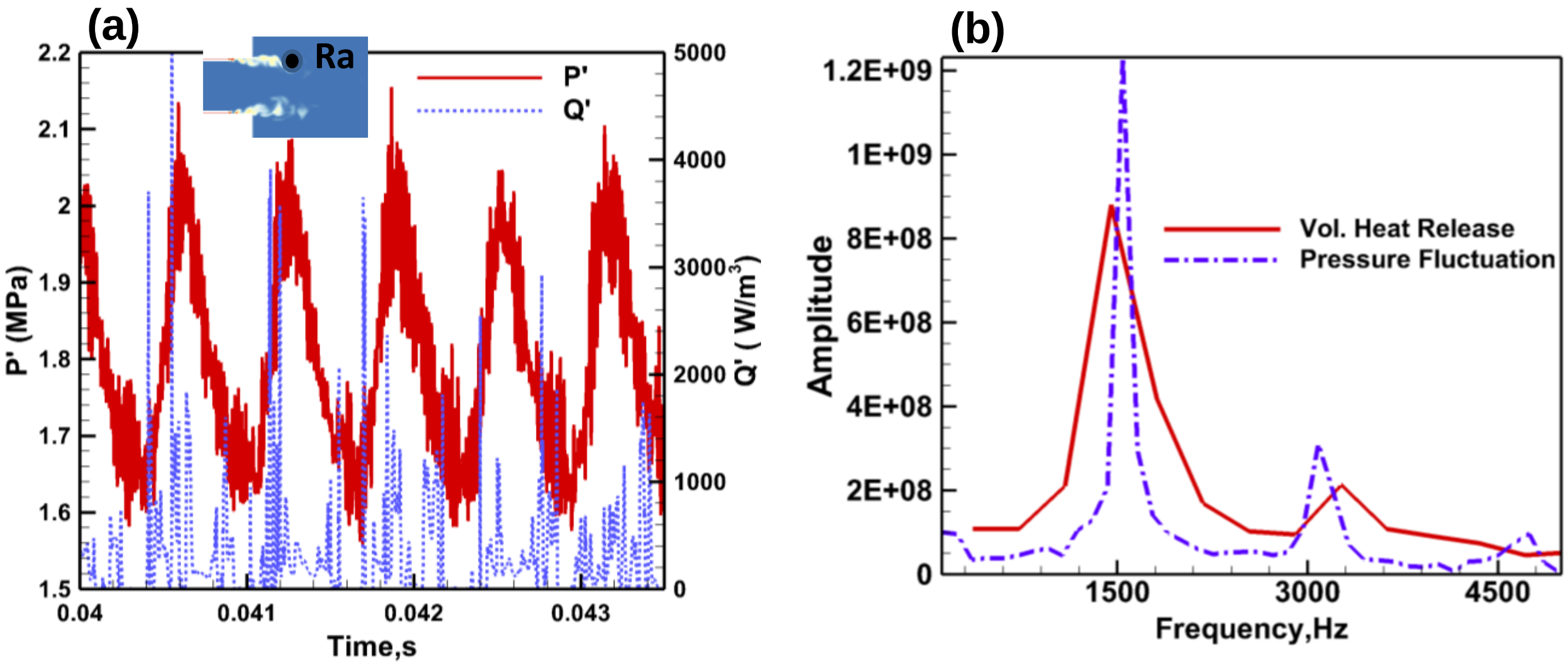

**FIG.13.**: (a) Heat release-pressure variation, and (b) FFT spectrum

The coupling of heat release and the acoustic wave is examined, displaying the effect of the traveling pressure wave on heat release. In Figure 13(a), a peak in heat release is observed with increased pressure at the head-end probe location. Higher heat release occurs upon the sudden disruption of vortices with unreacted oxidizer and accumulated fuel by the incoming compression wave at the backstep region. The increase in heat release due to the pressure wave, and vice versa, is clearly captured in the current LES. The FFT spectrum shown in Figure 13(b) reveals a match in the peak frequency of heat release and pressure. It demonstrates a positive feedback loop between pressure and acoustics, which sustains instability.

## E. Injector coupling hypothesis in CVRC

We further explore the instability mechanism and investigate the potential coupling of injector-chamber frequencies as the triggering factor for instability in CVRC. We discuss the pulse time theory presented by Harvazinski et al.[11] and extend it by including oxidizer-chamber frequency coupling. Figure 14 presents pressure variation in the oxidizer post, which shows out-of-phase pressure variation at probe locations POX-30 and POX-120, depicting a longitudinal wave movement in the post. POX-60 and POX-90 display similar pressure variations at the oxidizer post's center portion.

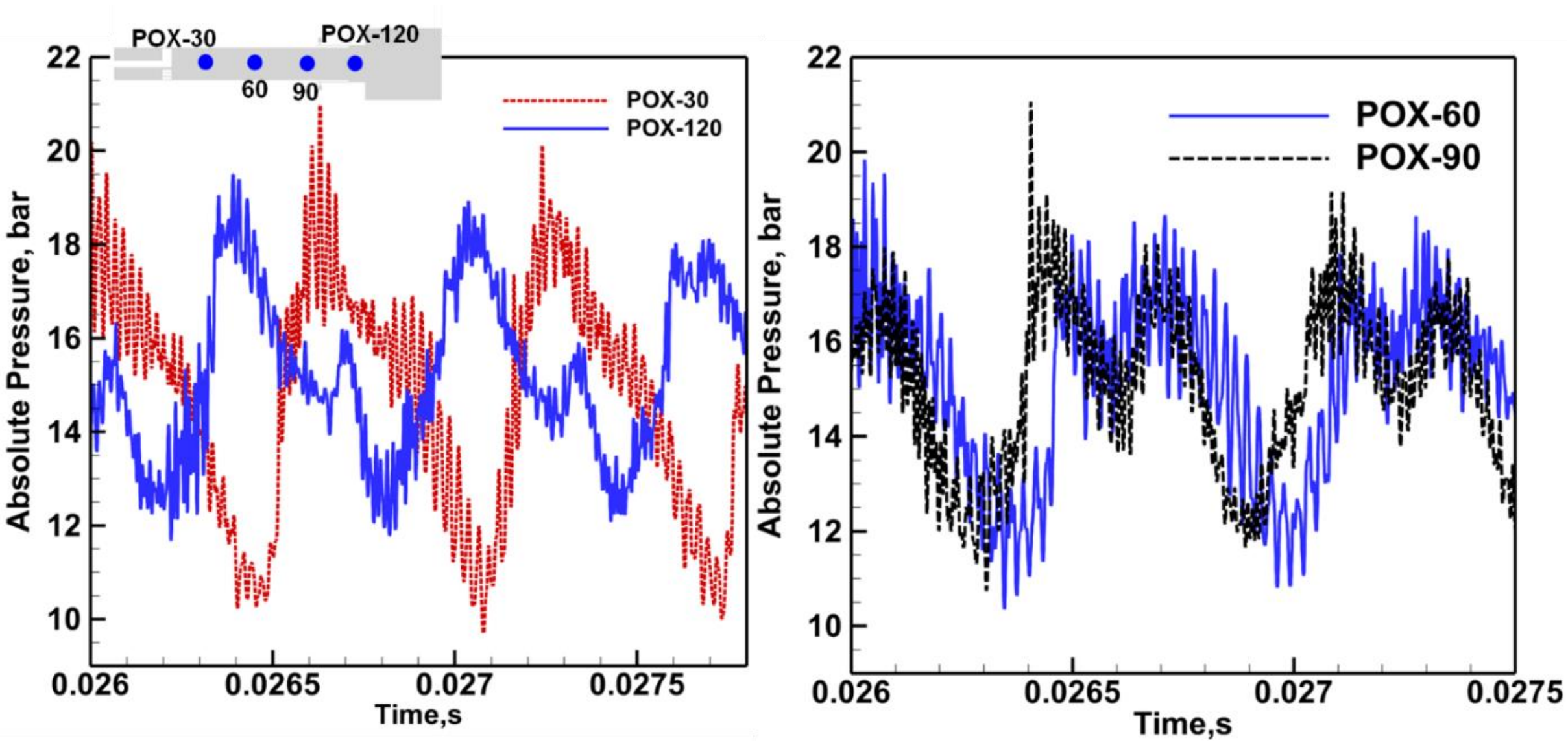

**FIG.14.**: Pressure variation in the oxidizer post

This pressure variation depicts the presence of acoustic wave movement in the post as well. A comparison of combustor (P1) and oxidizer post pressure (POX-120) is shown in Figure 15(a). It displays the same variation at both probe locations, indicating a similar acoustic wave characteristic. We perform an FFT of the oxidizer pressure spectrum. Figure 15(b) compares the peak frequency of the oxidizer and combustor probe, which shows a close match of frequency, with a dominant peak at 1520 Hz and corresponding harmonics. All the probes in the oxidizer post show the same frequency of 1520 Hz, the same as the first longitudinal mode of the combustor. The same frequency in combustor and oxidizer post locations indicates a frequency-coupled system, which is further elaborated.

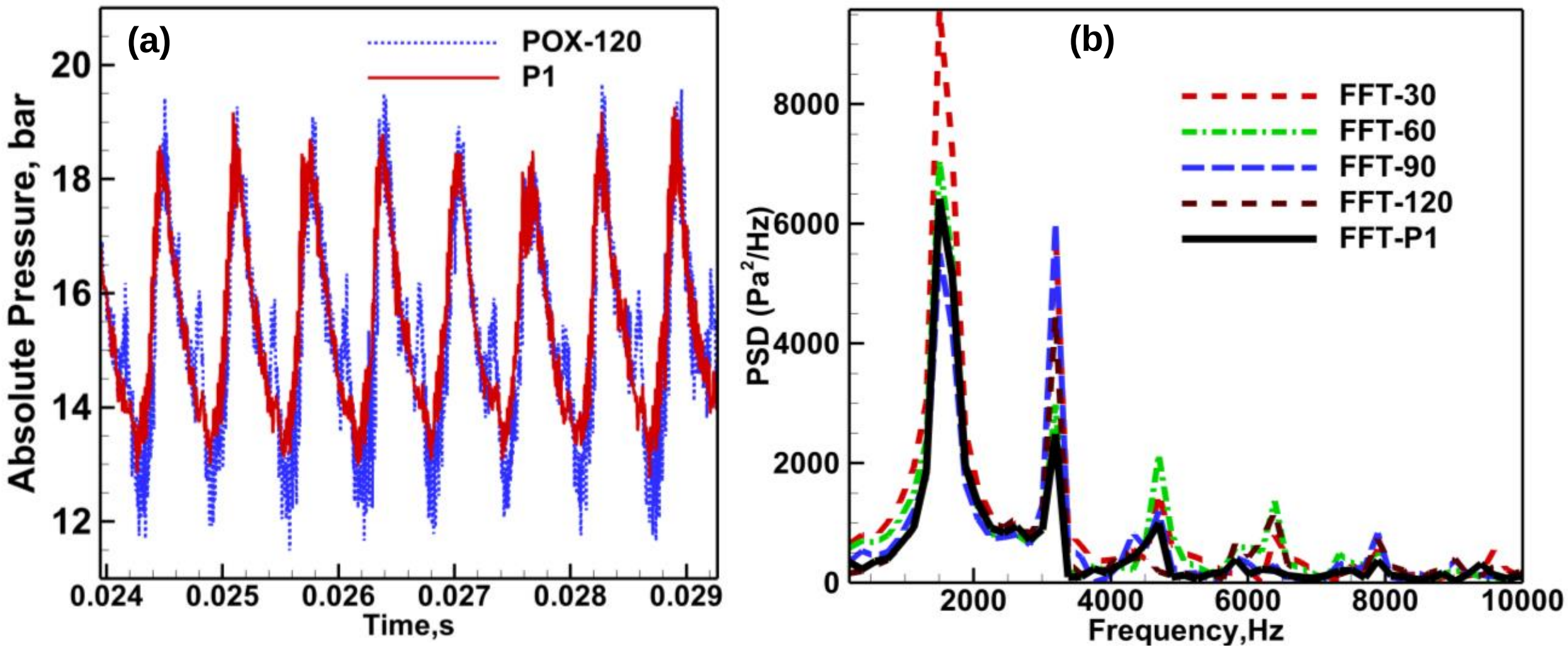

**FIG.15.**: (a) Pressure comparison in the oxidizer post and chamber, (b)FFT spectrum of oxidizer and chamber pressure data

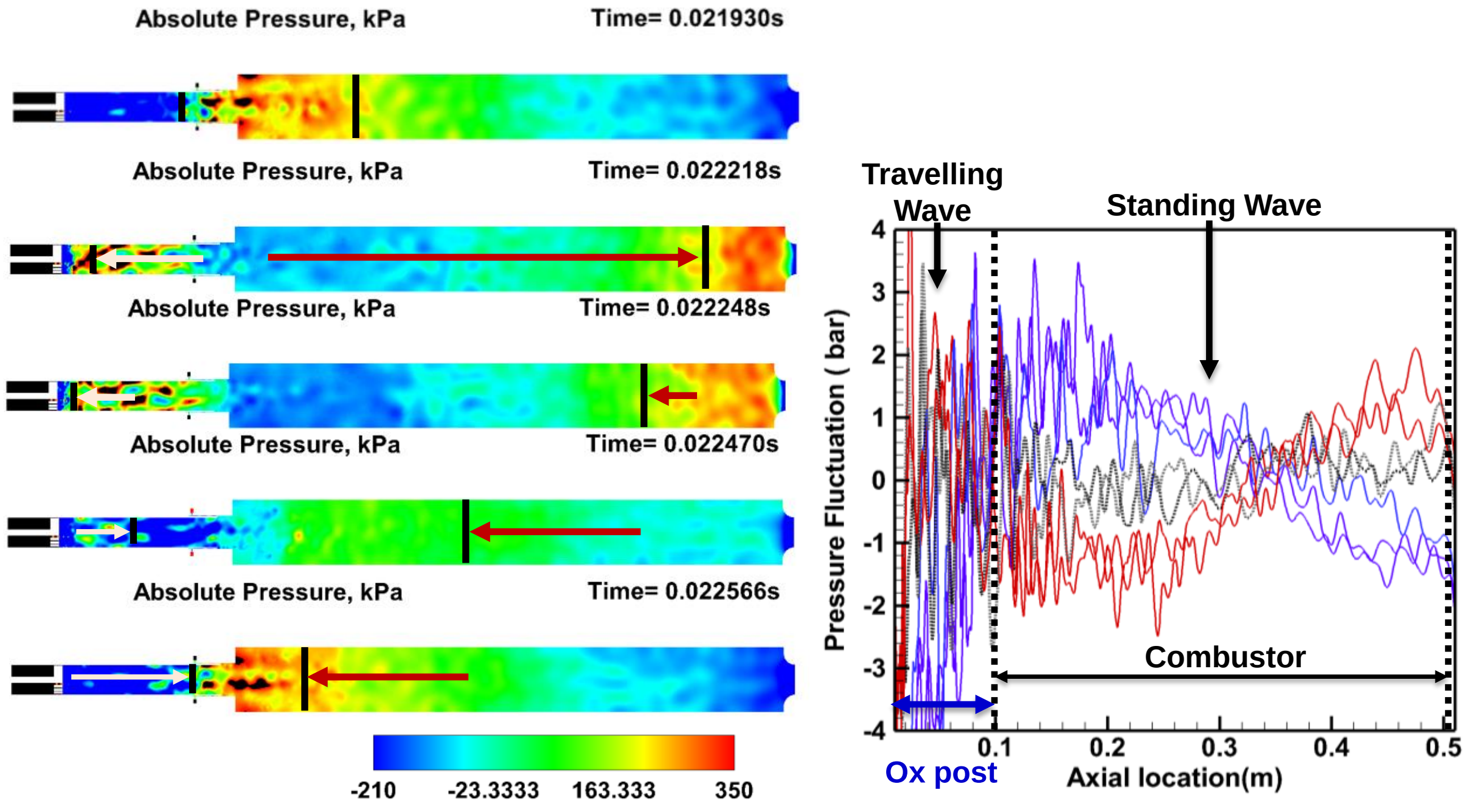

**FIG.16.**: (a)Acoustic wave movement in the oxidizer post and combustor, (b) Acoustic pressure variation at the centreline

Through analytical analysis, we investigate the frequency match-up or coupling of oxidizer post resonance with chamber acoustics. Considering the open-close acoustic boundary condition of the oxidizer post, the first longitudinal (1L) mode frequency of the oxidizer pipe ($C/4*L_o$) comes out to 1520 Hz at oxidizer post length, $L_o$: 0.12m and sound speed of C:730 m/s. A similar calculation for the combustor also provides longitudinal resonant mode ($C/2*L_c$) frequency of 1520Hz at chamber length, $L_c$:0.381m, and at chamber sound speed of C: 1150m/s considering close-close boundary conditions. The different acoustic boundary conditions for the oxidizer post and chamber reveal a match in resonant frequency, which can change the pulse time as proposed earlier in reference[11]. The oxidizer post-chamber frequency match-up can lead to a self-excited state. We observed that such a frequency match could result in methane accumulation in the backstep region, attributed to the return of the oxidizer pressure wave and lower head-end pressure. The frequency match eventually causes both the oxidizer and combustor waves to return to the head end simultaneously, leading to violent combustion and sustenance of instability in CVRC. The acoustic wave movement is pictorially represented in Figure 16 (a). It displays a longitudinal pressure wave pattern in the chamber and oxidizer tube in a cycle. It depicts the wave movement in the combustor section and oxidizer tube, simultaneously returning to the head end. The frequency coupling manifests in acoustic waves' simultaneous return, leading to sustained instability in this specific oxidizer post-length scenario. Figure 16(b) shows the pressure wave in the combustion chamber is more like a standing wave, although traveling effects are present. The wave inside the oxidizer post is more like a traveling wave. A coupling of traveling pressure oscillations inside the oxidizer post and pressure waves in the combustor model plays a crucial role in the instability.

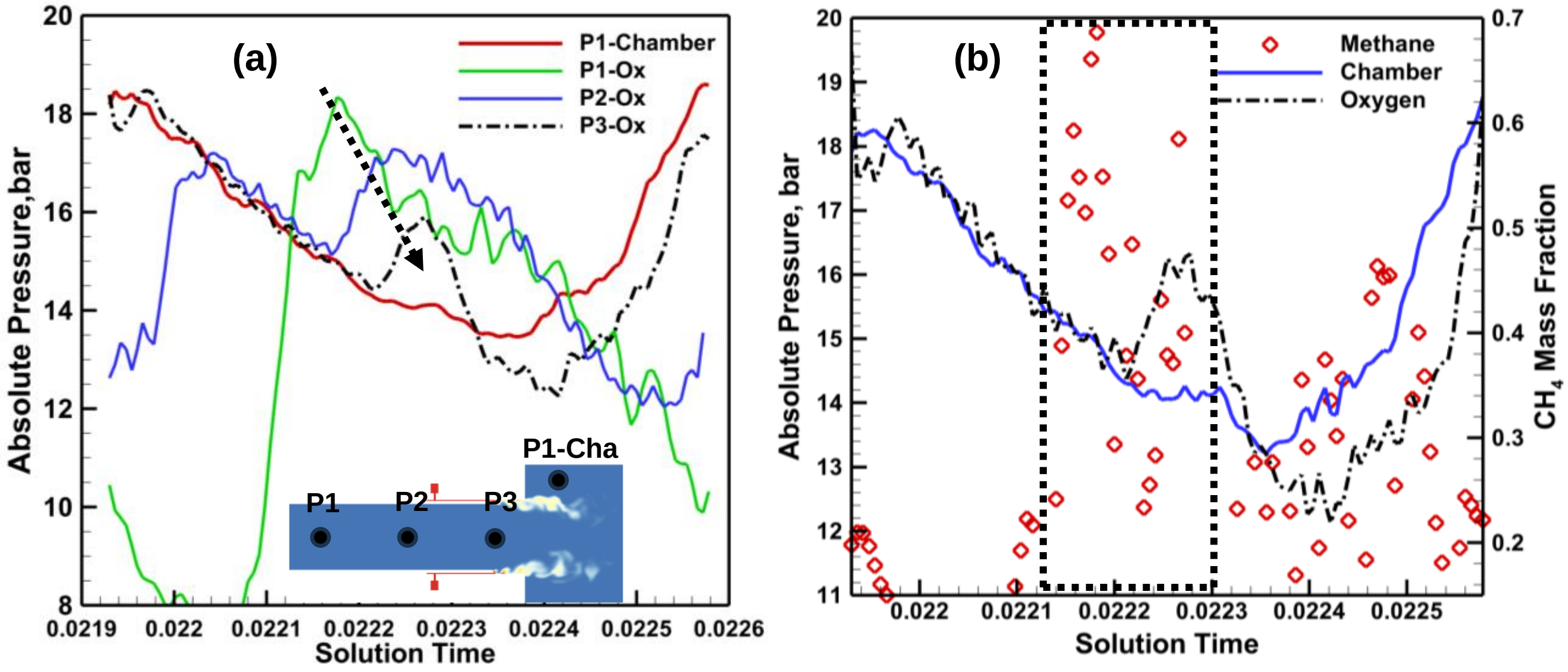

**FIG.17.**: (a) Acoustic pressure variation in oxidizer post and combustor, (b) Methane mass fraction evolution with oxidizer and combustor pressure

Figure 17(a) displays pressure variation at probes in the oxidizer post and the combustor. The P3 probe in the oxidizer post and P1-chamber probe follow a similar variation, with higher pressure at the end of one cycle. A sudden increase in oxidizer post pressure is attributed to the returning wave in the oxidizer post. The returning pressure wave can push methane into the lower head end pressure in the chamber. Figure 17(b) illustrates the sudden increase in methane mass fraction, with a jump in oxidizer pressure. This analysis showcases that fuel accumulates in the backstep region due to lower head end pressure and returning oxidizer tube pressure wave. The detailed analysis of the CVRC test case indicates that combustion instability arises primarily from the same resonant frequency in the oxidizer post and chamber for the specific post length of 12cm. The self-sustained instability in the CVRC is attributed to a combined effect of higher methane accumulation in the back step region, the simultaneous return of oxidizer and combustor pressure waves to the head end, and increased vortex shedding.

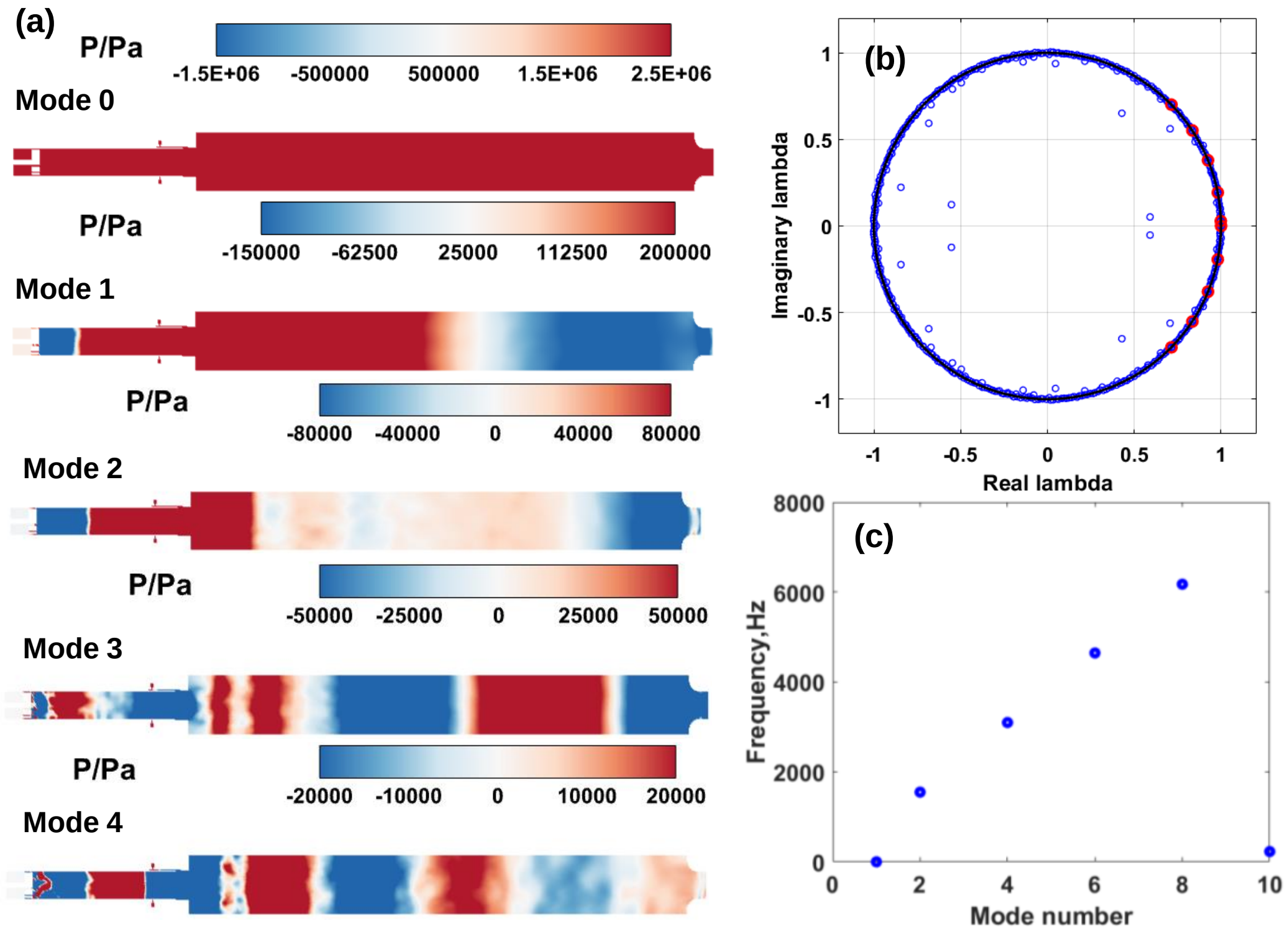

**FIG.18.**: (a) Mode shapes in CVRC (b) DMD Eigenvalues on the unit circle (c) Mode number vs. frequency plot

## F. Dynamic mode decomposition analysis: CVRC

To examine the instability modes in CVRC, we use dynamic mode decomposition. For the DMD analysis in this work, we have used 400 pressure snapshots taken at the axial plane, equivalent to more than 10 acoustic cycles. The pressure slice data is gathered at a sampling rate of 50 kHz. The spatial mode shapes are shown in Figure 18(a), where modes 1 to 4 represent 1L to 4L acoustic mode shapes, and mode 0 indicates the time-averaged characteristic of the flow field. Figure 18(b) displays the distribution of the DMD eigenvalues on the unit circle. The modes can be represented by a single mode with a positive frequency and occur in complex conjugate eigenvalue pairs.

Figure 18(c) shows the dominant frequency modes. It shows a harmonic behavior in the spectrum with the first mode at 1520 Hz, as captured in the FFT of pressure data. DMD illustrates acoustic waves' movement inside the combustor and injector sections. The acoustic modes reveal a potential acoustic resonance between the combustor and injector.

The case study shows the onset and propagation of the self-excited acoustic wave in the combustor, representing a standing wave-like structure corresponding to longitudinal resonant modes of the combustor. Using the LES methodology, the case study also serves as a validation study to simulate limit cycle self-sustained oscillations. An exact geometric model is generated with a choked oxidizer slotted inlet plunger and 36 inlet orifices for fuel entry into a high aspect ratio combustor ending with a choked throat. A combustor case with 12cm oxidizer post length is simulated, which displayed maximum PSD at first resonant longitudinal mode in experiments. Interaction of the acoustic wave with heat release is captured in the LES, with the variation of heat release at the pressure wave arrival and departure at the back step region at appropriate time instants. This case reveals the potential coupling of injector-chamber frequency, which, combined with hydrodynamic effects and similar pulse time movement, has enhanced combustion dynamics in CVRC. In the subsequent case study, we employ the LES and DMD methodology to investigate dynamics in a multi-element rocket-scale combustor.

## V. MULTI-ELEMENT COMBUSTOR CASE

This study utilizes a multi-element combustor case operating on a GOx-methane propellant combination. The primary objective is to simulate combustion dynamics during ignition transients of a full-scale rocket engine at representative sub-scale conditions. A multiple-injector configuration that simulates inter-element flow dynamics is employed in this case. Injectors serve as sensitive elements capable of generating and modifying flow oscillations. Understanding the dynamic characteristics of injectors is imperative for identifying their role in the performance and dynamics of rocket-scale combustors. This case study focuses on comprehending the influence of injector dynamics on combustion and capturing the possibility of injector-chamber acoustic coupling in a sub-scale rocket combustor. In this case study, we utilize a representative combustor with 7 bi-directional swirl injectors, with inter-element geometric features kept the same as the full-scale engine. The combustor is designed to operate at typical engine ignition conditions close to 10 bar, with methane and oxygen mass flow rate per element maintained the same as full-scale engine operation. The following section provides the configurational details of this test case.

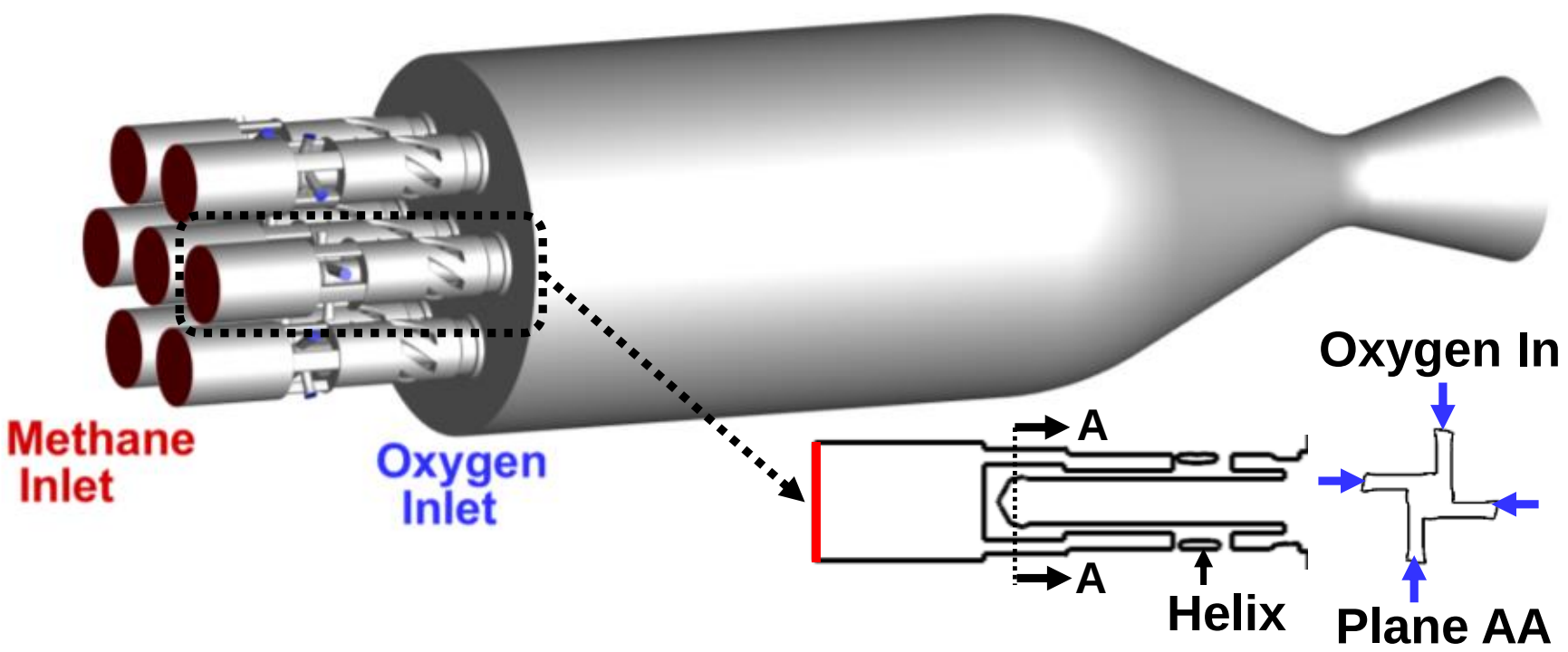

**FIG.19.**: Computational domain- Multi-Element Combustor & injector details

## A. Computational configuration of multi-element case

This section outlines the computational domain and input conditions for multi-element LES. The computational domain replicates a typical rocket engine with multiple injectors. Figure 19 illustrates the computational domain used in this case study. The injector elements feature a bi-directional swirl, coaxial, with an oxygen stream at the center and methane flow at the periphery. The propellants enter the combustor as separate non-premixed streams and exhibit rapid turbulent mixing at the injector exit region. The injector configuration is derived from the propulsion system successfully operating on LOx-hydrogen propellants. A bi-directional swirl is provided by tangential entry to the oxygen and helical vanes in the axial passage of methane to acquire swirl motion. The oxygen path axial length is $11.3R_n$, and the coaxial methane flow path length is $12.25R_n$, where $R_n$ is the oxygen post radius. A positive recess of $0.6R_n$ is provided to the oxygen post to enhance mixing. In this study, LES is performed under flow conditions that closely resemble the ignition conditions of the engine. This specific scenario is chosen to gain insight into combustion dynamics during the transient phase of engine operation. The selected condition allows for analyzing complex multi-element conditions within a reactive flow LES framework without invoking the real gas equation of state. The simulation uses a mass flow rate corresponding to the ignition condition, maintaining an overall oxidizer-to-fuel mixture ratio close to 2, with a mass flow rate of 30 gps for methane and 60gps for oxygen, respectively. Methane enters the injector element in the axial direction and traverses helical vanes to acquire swirling motion, while oxygen receives the tangential entry. Mass flow boundary conditions are applied at the methane and oxygen injector inlets. A pressure outlet boundary is imposed at the divergent outlet. The walls are treated with adiabatic conditions and a no-slip boundary. The methane injection condition is maintained at approximately 12 bar and 300 K, while the oxygen injection parameters are around 13 bar and 300 K. The corresponding injection density is 7 and 16 kg/m$^3$ for methane and oxygen, respectively. A significant difference in sound speed is noticed in methane (446 m/s) and oxygen (330 m/s) streams, which can impact the interaction of chamber acoustic waves with upstream injectors.

We utilize the mesh methodology as developed in our previous work[8]. A suitably refined mesh around the injector outlet region is created to resolve more than 80% of turbulent scales. The mesh quality aligns with the specifications outlined in the reference[8], which ensures a good quality LES mesh near the injector exit region to capture higher gradients. The resolved turbulent kinetic energy (TKE) ratio to total TKE gives an overall value above 0.8. The requirement of a lower acoustic CFL number to accurately capture pressure oscillations is maintained in this case study. Our previous research[8,13,50] suggests that an ACFL value of 6 is suitable for overall instability. In this case study, acoustic CFL is kept below 6, which determines the timestep size in this LES. The timestep size is adequate for capturing the unsteady dynamics originating in the multi-element chamber. All governing equations are implicitly filtered by the finite-volume methodology of ANSYS Fluent[49]. Spatial discretization uses second-order bounded schemes, while time integration is carried out with a bounded second-order implicit method. LES is conducted for many acoustic cycles to ensure independent results for initial conditions and to gather unsteady pressure statistics for spectral analysis. Our study presents spectral and dynamic mode decomposition analyses on pressure probes and spatial data. The study describes the interplay between the injector and chamber on the enhanced combustion dynamics in this multi-element test case. We discuss the instantaneous flame features before spectral and DMD analysis.

## G. Instantaneous flow features

We discuss the instantaneous flow and flame characteristics before examining pressure probe data and performing spectral analysis to determine dominant frequency modes in the chamber and injector regions. Pressure data is collected from several probe positions in the chamber and injector throughout over 35 acoustic cycles of LES. At a time of 0.04656s, we emphasize essential immediate characteristics in contour plots. The instantaneous pressure variation over the combustor at the axial and radial planes is shown in Figure 20. The pressure variation is displayed independently in the radial planes 1, 2, and 3. High-pressure areas are shown by pressure variation in an axial plane, which suggests the existence of a high-frequency acoustic wave. The injector segment's staggered pressure variation shows the acoustic wave's travel in the injectors. Planes 1, 2, and 3 display radial pressure variation, representing a typical high-frequency transverse acoustic wave in the combustor. Figure 21 displays the instantaneous flame temperature at an axial plane. The flame contour describes rapid turbulent mixing and combustion in the downstream region of the injector exit. It exhibits an anchored flame at the oxygen post and the mixing of high-temperature turbulent eddies along the combustor axis. The flame temperature dictates the sound speed in the chamber region, where a higher sound speed is noticed, whereas the oxygen path of injectors exhibits a lower sound speed of 300 m/s. A drastic change in sound speed is observed, which can impact the resonant modes of the combustor. Figure 22 presents the vorticity magnitude at an axial plane, displaying a high vorticity region in the core and exit regions of the injectors. The high vorticity region is attributed to higher velocity gradients associated with the oxygen swirl at the injector core. The high-temperature flame region at the injector exit induces high-velocity gradients, generating higher vorticity. Instantaneous LES contours showcase physical flow and flame features, while spectral and dynamic mode decomposition analysis is followed to explore injector-chamber frequency-coupled combustion dynamics.

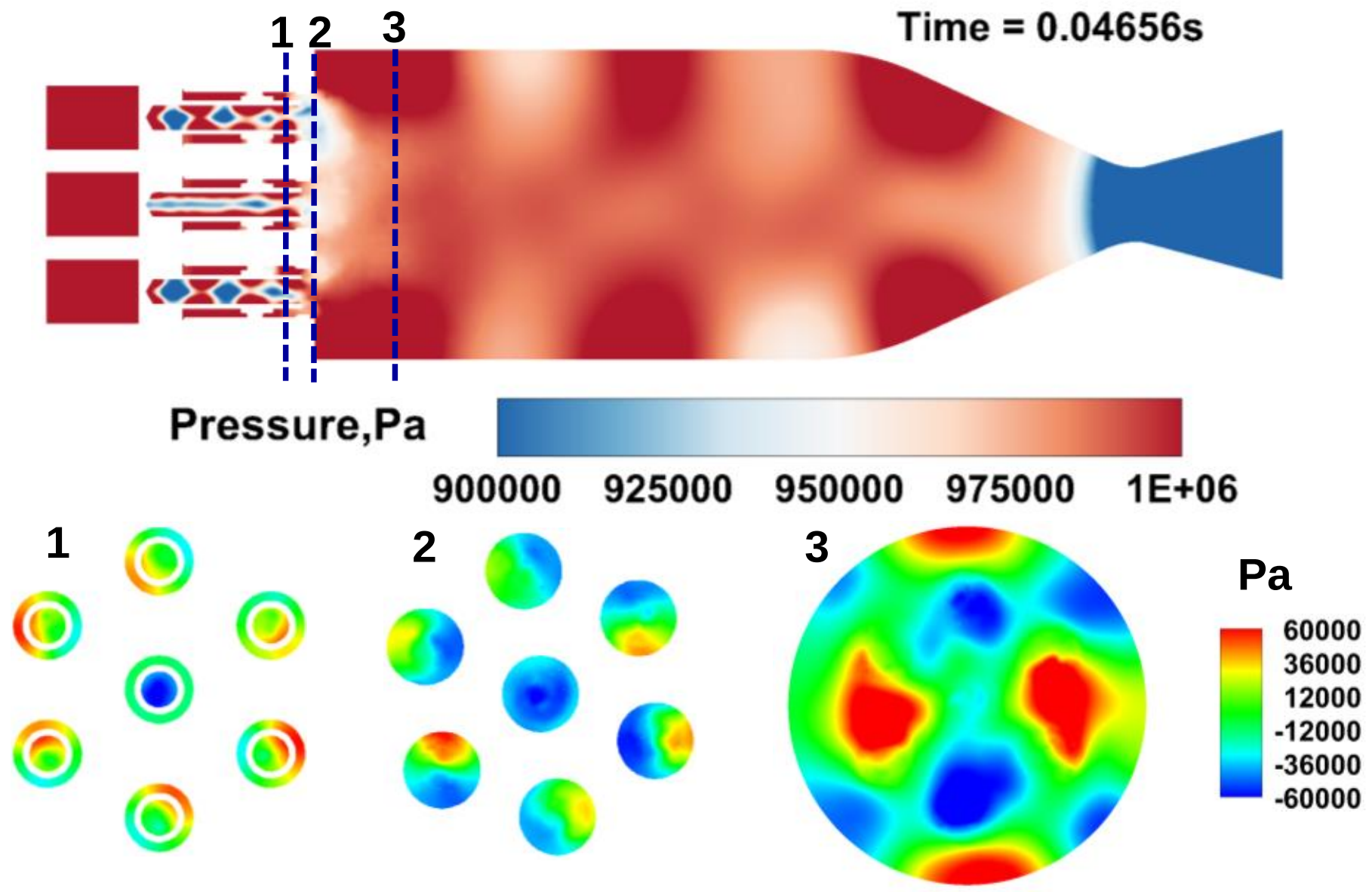

**FIG.20.**: Instantaneous pressure variation at axial and radial planes

The assessment of combustion dynamics includes analyzing the evolution of fluctuating pressure waves within the combustor. Temporal changes in pressure provide insights into the dynamics of the combustor. Our initial case study illustrated the initiation of longitudinal pressure fluctuations in CVRC. In this case study, we gather absolute pressure data over various locations at a sampling frequency of 1 MHz. Figure 21 also displays the pressure probe locations at the chamber wall and in the injector region of the oxygen post. Pressure data is gathered in the injector region to unveil the mechanisms related to combustion dynamics associated with the injector in the combustor.

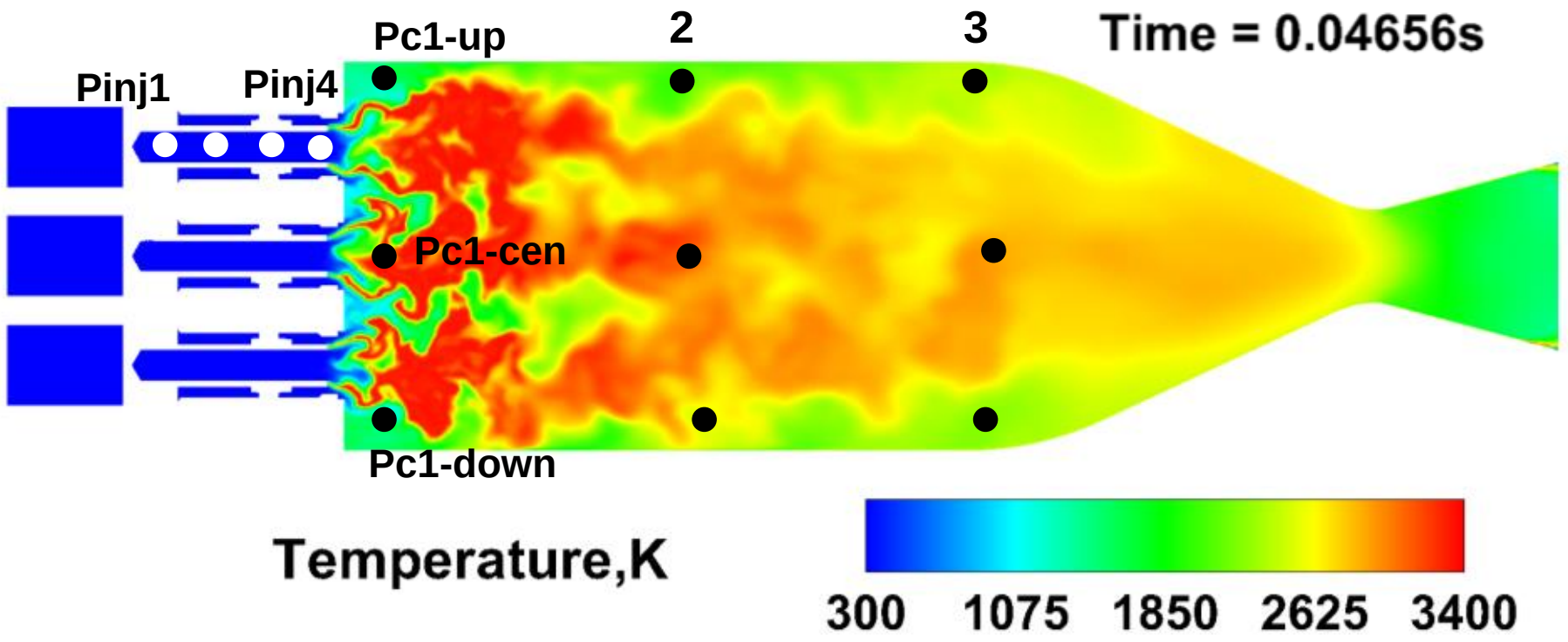

**FIG.21.**: Instantaneous flame temperature at an axial plane

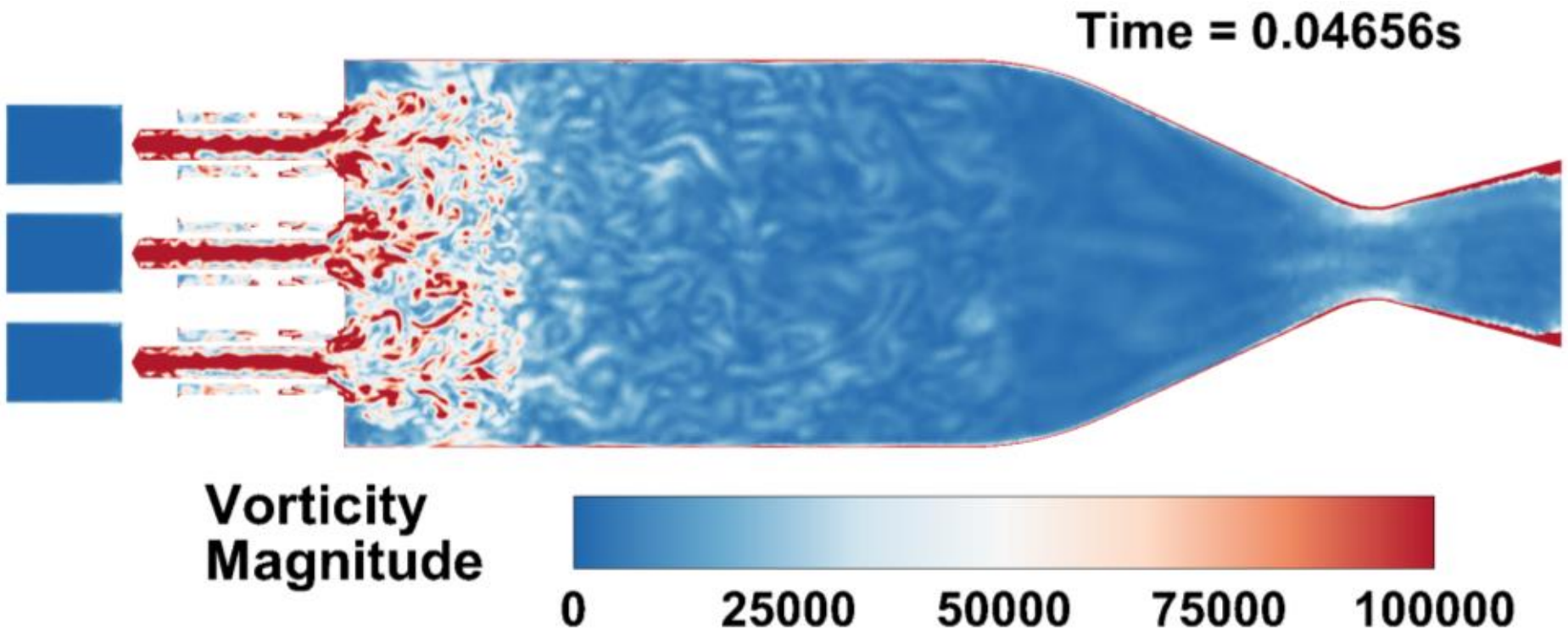

**FIG.22.**: Instantaneous vorticity magnitude at an axial plane

Figures 23 (a) and (b) show raw absolute pressure time histories in the chamber and injector. Figure 23(a) displays the peak-to-peak pressure variation above the mean pressure of 10 bar at probe locations: up (Pc-up), center (Pc-cen), and down (Pc-down) points. It shows higher fluctuating pressure at the near wall probes (Pc-up & Pc-down) while displaying lower amplitude at the center point. The circumferentially opposite probes display an out-of-phase pressure variation with anti-nodes near the wall. The pressure variation indicates the movement of transverse acoustic waves in the combustor. These pressure wave characteristics are extracted through spectral analysis of the pressure probe data and presented in the subsequent section. Figure 23 (b) displays pressure variation plotted at 4 marked points inside the injector section. It reveals an out-of-phase pressure variation for the injector points Pinj1 and Pinj2, as well as for Pinj3 and Pinj4, respectively. The pressure variation demonstrates a typical longitudinal wave pattern in the oxygen path. The oxygen path manifests a higher mode longitudinal wave, even though it has the potential to exhibit transverse modes. However, the probes do not capture these transverse modes due to their longitudinal placement.

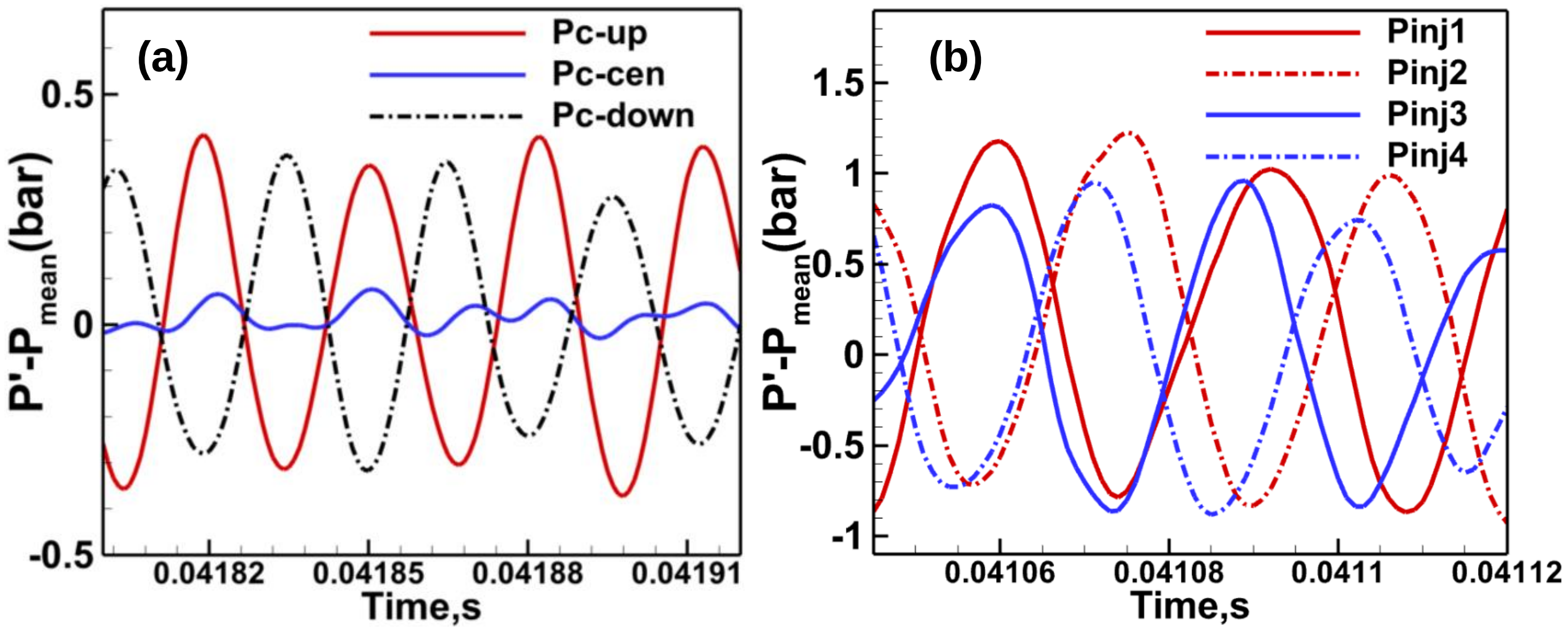

**FIG.23.**: Pressure variation in (a) chamber and (b) injector probe locations

Figure 24 presents a spectral plot for chamber points. In Figure 24(a), a high amplitude discrete tone close to 32 kHz is observed for all chamber points near the chamber wall. A closer examination of the frequency values reveals a peak at 4 kHz and subsequent peaks at 10, 18, and 24 kHz, respectively. Figure 24(b) illustrates a 4 kHz frequency in probes Pc1-up and Pc3-up, which is not visible at the Pc2-up probe location. Spectral analysis identifies the 4 kHz frequency as a longitudinal mode of the combustor.

Analytical calculations for resonant modes also confirm the first longitudinal mode at 4 kHz and the first tangential mode frequency at 10 kHz. The dominant peak captured at 32 kHz represents the higher harmonics of the first tangential mode, as identified through pressure variation in Figure 23(a). Figure 24(c) further illustrates the presence of a transverse wave in the combustor, where near-wall probes Pc2-up and Pc2-down exhibit a dominant peak at 32 kHz, while the Pc2-cen probe FFT shows no peak at 32 kHz. A similar variation is observed at the Pc3 sensor location, exhibiting a peak for near-wall probes and no peak for the center probe (Pc3-cen). This confirms the onset of a high-frequency transverse wave in the combustor, potentially impacting the upstream injectors. In Figure 25, we compare the FFT of the chamber and injector probe data. It closely matches the FFT peak in the injector and chamber section. It shows the frequency for all injector probes in a close match with chamber probes. The spectral plot depicts pressure dynamics similar to those in the injector and chamber but with more violent dynamics in the injector than in the chamber region. FFT comparison demonstrates a close match of frequency (32kHz) in the injector and chamber section, indicating a high chance of acoustic resonance between the injector and chamber. The injector dynamics can affect the overall stability of the combustor, which manifests from injector-chamber frequency coupling. This frequency coupling can lead to fluctuations in injector mass flow rate, eventually affecting the stability of the combustor. This case study highlights the possibility of acoustic coupling between the combustor's transverse mode and longitudinal mode of oxygen path. The oxygen path, in resonance with the combustor, can act as a source of fluctuations. A frequency coupling between the longitudinal mode oscillations in the oxygen post and the transverse mode of the combustor can disrupt the oxygen flow periodically, leading to oscillations in the mass flow rate. It contributes to unstable heat release and consequent pressure fluctuations in the chamber. We noticed similar instability driven in the experimental BKD combustor[6,47] due to injector-chamber frequency coupling. A similar coupling phenomenon has been recently reported through numerical investigation in references[28,29]. Our case study reveals that such frequency match-up can potentially regulate the combustion dynamic process in the combustor.

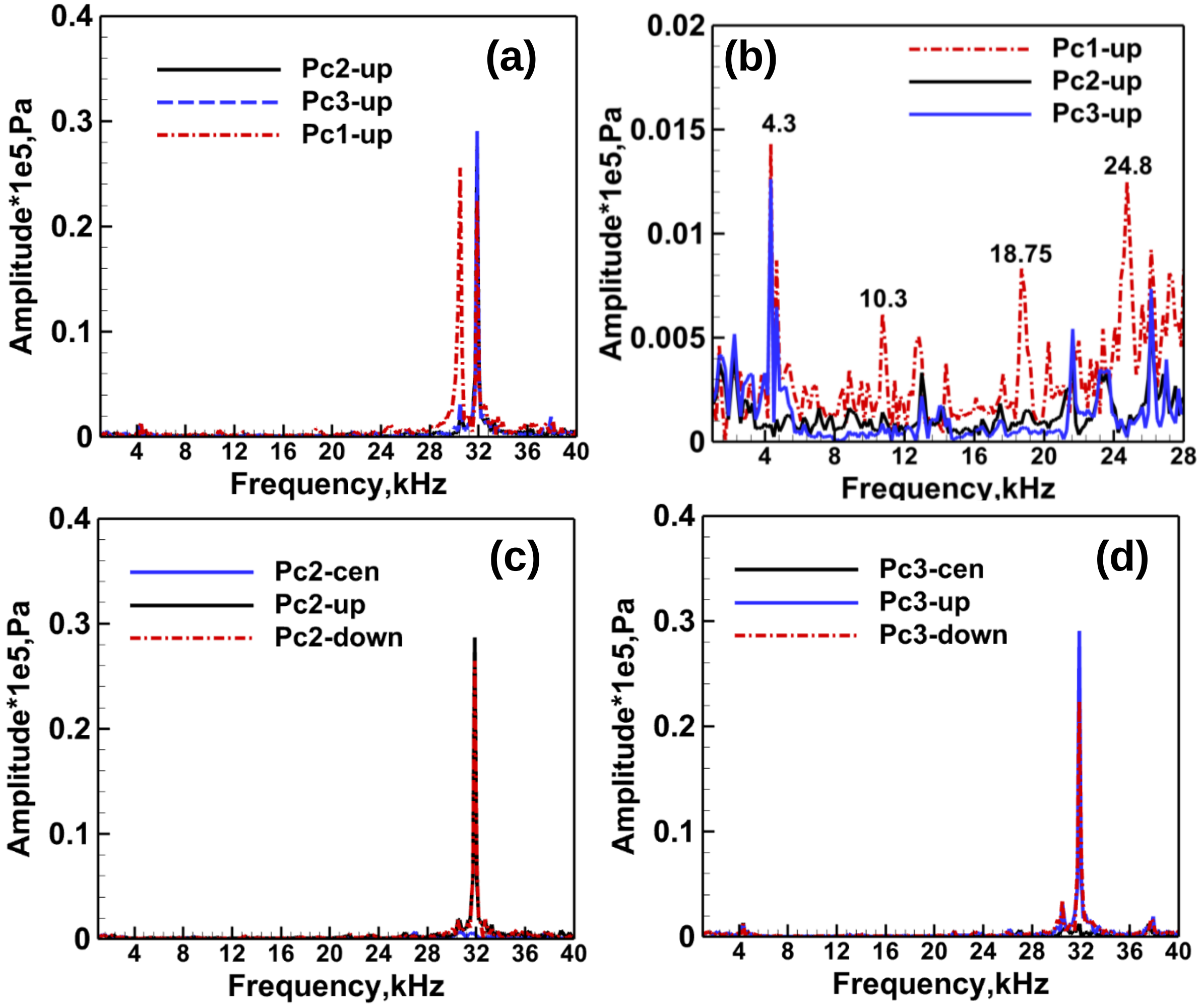

**FIG.24.**: Spectral plots pressure: (a) near champer wall, (b) near dump plane, (c) middle of chamber, (d) end of chamber

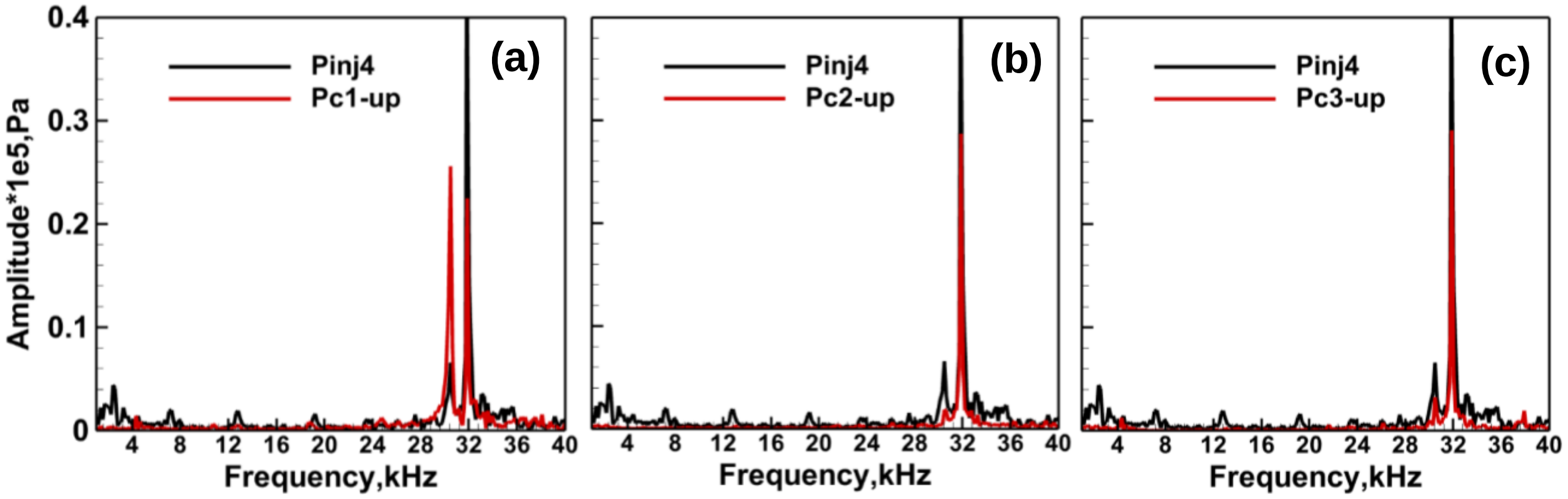

**FIG.25.**: Frequency comparison between chamber and injector probe: (a) near dump plane, (b) middle of the chamber, (c) end of chamber

## H. Dynamic mode decomposition analysis: Multi-element combustor

We have also performed dynamic mode decomposition to examine the instability modes in a multi-element combustor for this case. In this case study, we utilized 300 snapshots of pressure at the axial plane, corresponding to more than 20 acoustic cycles, for DMD analysis gathered at a sampling rate of 100 kHz. Figure 26 displays spatial mode shapes, where mode 0 denotes the time-averaged characteristic of the flow field. Meanwhile, modes 2 onwards display a higher transverse mode acoustic setup in the chamber, as observed in spectral analysis. The spatial modes in the chamber display a typical representation of high-frequency transverse modes, whereas a longitudinal acoustic standing wave is visible in the injector section. These structures are evident in all spatial modes, as shown in Figure 26. An enlarged injector region is shown in Figure 26, which illustrates the presence of longitudinal acoustic waves in the injector region.

The DMD analysis could capture the movement of acoustic waves into the injector region. It reveals the potential of acoustic waves in the chamber to propagate upstream into the injectors. Specifically, the analysis demonstrates that the acoustic wave movement inside the chamber (transverse modes) excites higher longitudinal modes within the injectors near the wall. It should be noted that transverse acoustic waves do not influence the central injector in the chamber. Such an uneven response of upstream injectors can lead to mass flow rate fluctuations and can cause enhanced combustion dynamics. The acoustic modes also reveal a potential acoustic resonance between the combustor and injector, which can modulate the injector flow rate corresponding to the dominant mode of the combustor. DMD analysis of multi-element combustors also showcases upstream injectors' interdependence on chamber acoustics, with potential injector-chamber frequency coupling leading to elevated combustion dynamics.

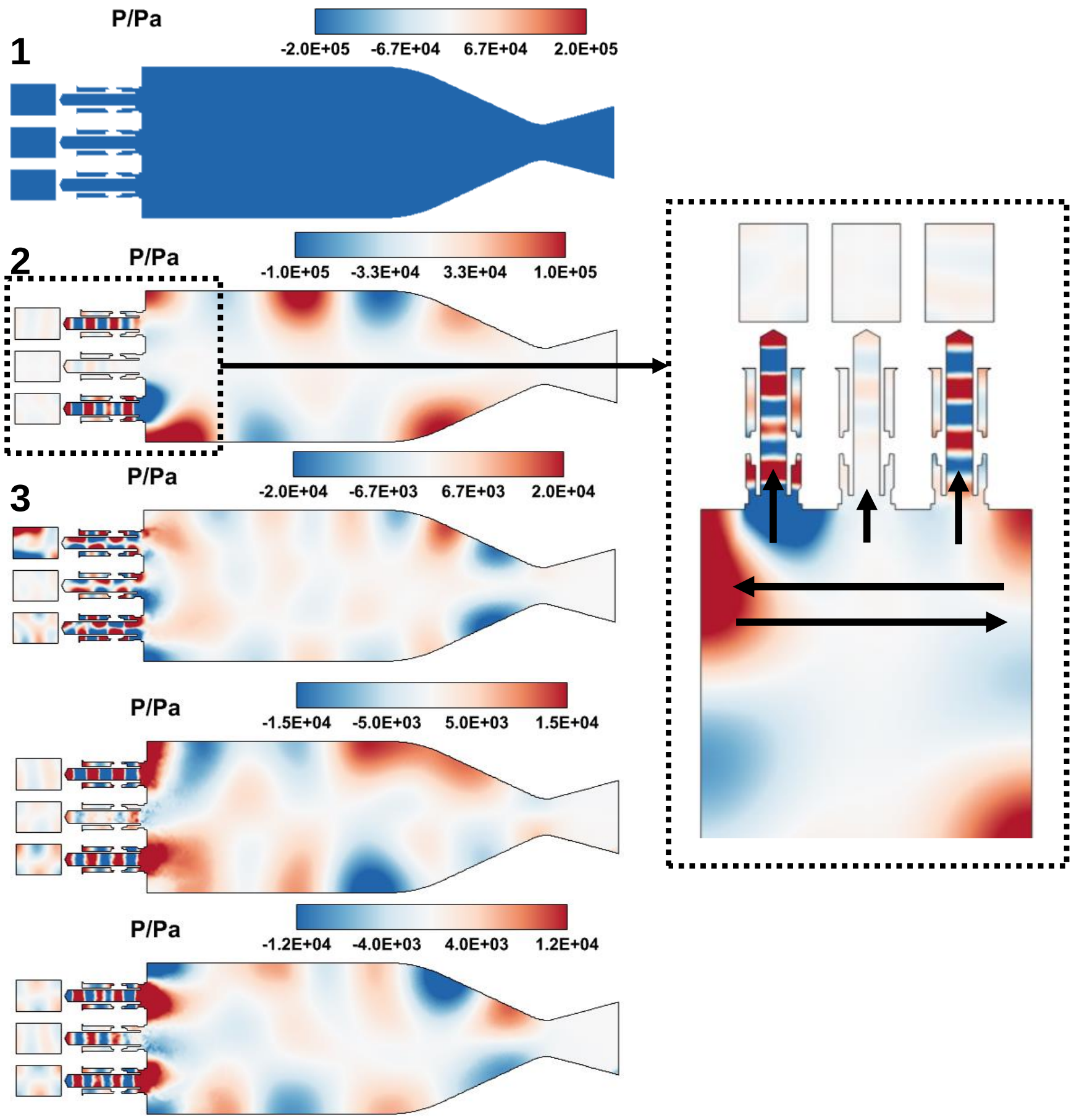

**FIG.26.**: DMD spatial mode shapes in a multi-element combustor with zoomed injector section

## VI. CONCLUSION

First, we investigate the benchmark example of CVRC using LES modeling and decomposition analysis. A thorough analysis provides a clear understanding of the instability mechanism in CVRC, emphasizing the possibility of injector-chamber frequency coupling. In CVRC, an unstable thermo-acoustic state can result from a frequency match in a combustor and oxidizer post length of 12 cm. The spatial acoustic modes in CVRC reveal injector-combustor frequency coupling. Our second case study focuses on a multi-element combustor to capture the dynamic aspects of combustion close to ignition conditions. The spectral analysis identifies pressure oscillations corresponding to the injectors' longitudinal and the chamber's transverse modes. Similar acoustic frequencies are detected by the spectrum analysis in the vicinity of the injector and combustor, suggesting that frequency coupling may be the cause of the combustion dynamics. The potential acoustic resonance between the chamber and injector can modulate the injector flow rate corresponding to the combustor's dominant mode. DMD analysis shows that higher longitudinal modes within the injectors close to the wall are excited by the transverse acoustic wave movement inside the chamber. However, the transverse wave mode does not affect acoustics in the central injector. Moreover, the variations in mass flow rate caused by an uneven response from upstream injectors can eventually result in enhanced combustion dynamics in the multi-element combustor configuration. This work demonstrates how the acoustic modes within the combustor impact the hydrodynamics of the injector and explains how the coupling between the injector and chamber frequencies leads to high-frequency instability.

## ACKNOWLEDGMENT

The LES case is executed on the High-performance computing facility established at LPSC using 560 direct water-cooled Intel Broadwell compute cores.

## AUTHOR DECLARATIONS

### CONFLICT OF INTEREST

The authors have no conflicts to disclose.

### DATA AVAILABILITY

The data that support the findings of this study are available from the corresponding author upon reasonable request.